\documentclass[twocolumn,prd,aps,longbibliography,superscriptaddress,preprintnumbers,nofootinbib,eqsecnum,amsfonts,amsmath]{revtex4-1}

\usepackage{graphicx}

\usepackage{amssymb}

\usepackage{amsmath}
\usepackage{subfigure}
\usepackage{verbatim}
\usepackage{physics}
\usepackage[colorlinks=true,allcolors=blue]{hyperref}
\usepackage[utf8]{inputenc}

\usepackage{color}

\usepackage{float}
\begin{document}
\title{Forecasts for detecting the gravitational-wave memory effect with Advanced LIGO and Virgo}

\author{Oliver M. Boersma}
\email{o.m.boersma@uva.nl}
\affiliation{Anton Pannekoek Institute for Astronomy, University of Amsterdam, Science Park 904, P.O. Box 94249,
1090 GE Amsterdam, The Netherlands}
\affiliation{Department of Astrophysics, Faculty of Science, Radboud University Nijmegen,
P.O. Box 9010, 6500 GL Nijmegen, The Netherlands}

\author{David A. Nichols}
\email{david.nichols@virginia.edu}
\affiliation{Department of Physics, University of Virginia, 382 McCormick Road, P.O.~Box 400714 Charlottesville, Virginia 22904-4714, USA}
\affiliation{Gravitation Astroparticle Physics Amsterdam (GRAPPA), University of Amsterdam, Science Park 904, P.O. Box 94485,
1090 GL Amsterdam, The Netherlands}
\affiliation{Department of Astrophysics, Faculty of Science, Radboud University Nijmegen,
P.O. Box 9010, 6500 GL Nijmegen, The Netherlands}

\author{Patricia Schmidt}
\email{pschmidt@star.sr.bham.ac.uk}
\affiliation{School of Physics and Astronomy and Institute for Gravitational Wave Astronomy,
University of Birmingham, Edgbaston B15 2TT, United Kingdom}
\affiliation{Department of Astrophysics, Faculty of Science, Radboud University Nijmegen, P.O. Box 9010, 6500 GL Nijmegen, The Netherlands}

\begin{abstract}
The detection of gravitational waves (GWs) from binary black holes 
(BBHs) has allowed the theory of general relativity to be tested in 
a previously unstudied regime: that of strong curvature and 
high GW luminosities.
One distinctive and measurable effect associated
with this aspect of the theory is the nonlinear GW memory effect. 
The GW memory effect is characterized by its effect on freely 
falling observers: the proper distance between their locations 
differs before and after a burst of GWs passes by their locations. 
Gravitational-wave interferometers, like the LIGO and Virgo 
detectors, can measure features of this effect from a single BBH
merger, but previous work has shown that it will require an event 
that is significantly more massive and closer than any previously
detected GW event.
Finding evidence for the GW memory effect within the entire
population of BBH mergers detected by LIGO and Virgo is more 
likely to occur sooner.
A prior study has shown that the GW memory effect could be detected
in a population of BBHs consisting of binaries like 
the first GW150914 event after roughly one-hundred events.
In this paper, we compute forecasts of the time it will take the
advanced LIGO and Virgo detectors (when the detectors are 
operating at their design sensitivities) to find evidence for
the GW memory effect in a population of BBHs that
is consistent with the measured population of events in the first
two observing runs of the LIGO detectors. 
We find that after five years of data collected by the advanced 
LIGO and Virgo detectors the signal-to-noise ratio for the 
nonlinear GW memory effect in the population will be about three
(near a previously used threshold for detection).
We point out that the different approximation methods used to
compute the GW memory effect can lead to notably different 
signal-to-noise ratios.
\end{abstract}

\maketitle

\section{Introduction}
\label{S:1}

The first detection of a gravitational waves (GWs) from a pair of 
merging black holes (GW150914) by the LIGO 
detectors~\cite{TheLIGOScientific:2016wfe}
opened a new avenue for testing the predictions of general relativity
for strongly gravitating and rapidly evolving 
spacetimes.
The observed GWs were consistent with the predictions 
of general relativity (GR) to within the statistical uncertainties 
of the measurement~\cite{TheLIGOScientific:2016src}. 
The LIGO, and subsequently Virgo, detectors have continued to discover
new GW events: after the first two observing runs of 
the two LIGO detectors now ten GWs from binary-black-hole (BBH) 
mergers and one from a binary-neutron-star merger have been
discovered~\cite{ligo2018gwtc}. 
(The GWs from these ten additional events are also consistent with
the predictions of GR~\cite{LIGOScientific:2019fpa}.)
Already in the third observing run of LIGO, over 20 BBH candidate 
events have been announced~\cite{GraceDB} and this number will 
rapidly increase once the detectors reach their design sensitivities
in a few years~\cite{LIGOScientific:2019vkc}.
The improved sensitivity of the detectors and the large number 
of events will allow GR to be tested more 
precisely for a range of binaries with different masses 
and spins.

Before the LIGO and Virgo discoveries, the predictions of GR were
consistent with a range of experiments and measurements in the  
Solar System and through observations of pulsars in the Milky Way
(see, e.g.,~\cite{Will:2014kxa} for a review).
Solar System, pulsar, and BBH observations probe different 
aspects of Einstein's theory: most importantly, BBHs 
allow GR to be studied in a more nonlinear and highly radiating 
regime of the theory than either Solar System experiments or pulsar
observations.  
Thus,  BBHs will allow the study of gravitational phenomena 
that require nonlinearities and high GW luminosities. 
One such effect, called the nonlinear GW memory 
effect~\cite{Christodoulou:1991cr,Blanchet:1992br},
is the focus of this paper.

The GW memory effect is characterized by a lasting change in the 
GW strain that occurs for many types of transient GW sources.
Zel'dovich and Polnarev~\cite{Zeldovich:1974gvh} first computed the 
GW memory effect in linearized gravity when they computed the GWs 
emitted by the gravitational scattering of compact 
objects.\footnote{Note that the possibility of the GW memory 
effect was considered by Newman and Penrose in~\cite{Newman:1966ub},
although they did not explicitly calculate the effect from any 
source.}
The high luminosities of neutrinos from supernovae also can produce
the GW memory effect, as was shown by Epstein~\cite{Epstein1978} and
Turner~\cite{Turner1978}.
Christodoulou~\cite{Christodoulou:1991cr} showed that there is 
also a nonlinear contribution to the effect in the full theory of GR 
(without the linear approximation), which arises from the energy flux
(luminosity per solid angle) from the GWs.
Blanchet and Damour~\cite{Blanchet:1992br} independently 
computed the effect within the context of the multipolar-expanded 
post-Minkowskian approximation. 
Binary black holes, with their high GW luminosities, are 
expected to have a non-negligible GW memory effect (see, 
e.g.,~\cite{Wiseman:1991ss,Favata:2008yd} for 
calculations in post-Newtonian theory 
and~\cite{Pollney:2010hs} for computations of the 
GW memory in numerical-relativity simulations).

The GW memory effect can be measured, because when a GW with 
memory passes by freely falling observers, the proper displacement
between the observers differs before and after the burst of GWs 
pass by their locations. 
The GW memory effect also has close connections to the symmetry 
group of asymptotically flat spacetimes, the Bondi-Metzner-Sachs 
group~\cite{Bondi1962,Sachs1962a,Sachs1962b}, and its corresponding 
conserved quantities (see, e.g.,~\cite{Strominger:2017zoo} for 
more details).
Thus, because of its distinctive observational signature and its
close connection to fundamental aspects of asymptotically flat
spacetimes, the GW memory effect would be of great interest to
detect.

The GW memory effect is formally the constant difference in the GW 
strain before and after a burst of GWs passes by a GW detector.
Interferometric GW detectors like LIGO and Virgo, however, are 
sensitive to GWs over a finite frequency range; thus, they do not
always have the necessary sensitivity at low frequencies to measure
the lasting change in the GW strain associated with the memory.
Nevertheless, the simulations in~\cite{Pollney:2010hs}
confirmed the analytical approximation used 
in~\cite{Favata:2009ii}, which showed that the memory effect rapidly
settles to a nonzero constant value over a timescale
(and hence frequency range) that LIGO and Virgo can measure,
for stellar-mass BBHs.
The prospects for measuring the GW memory effect from the 
full inspiral-merger-ringdown waveform of a BBH showed more 
promise for detecting the effect than earlier studies using
just the post-Newtonian approximation to the waveform during
the inspiral~\cite{Thorne:1992sdb,Kennefick:1994nw}.
However, Favata~\cite{Favata:2009ii} and more recently Johnson 
\textit{et al}.~\cite{Johnson:2018xly} showed that for LIGO and Virgo
to detect the GW memory from a single BBH merger would require 
a much closer or more massive BBH event than had 
previously been observed. 
Next-generation ground-based detectors such as the Einstein 
Telescope~\cite{Punturo:2010zz} and Cosmic 
Explorer~\cite{Reitze:2019iox} were shown
in~\cite{Johnson:2018xly} to be much more likely to detect 
the GW memory effect.
The planned space-based GW detector, LISA~\cite{Audley:2017drz} 
could detect the GW memory from supermassive BBH mergers (see, 
e.g.,~\cite{Favata:2009ii}). 
Pulsar timing arrays (see, e.g.,~\cite{Hobbs:2017zve}) have also
put constraints on GWs with memory (see~\cite{Aggarwal:2019ypr} 
and references therein), though there are forecasts that suggest 
pulsar timing arrays are less likely than LISA is to detect the GW memory effect~\cite{Islo:2019qht}.

Instead of searching for the GW memory effect associated with a
single BBH merger, Lasky \textit{et al}.~\cite{Lasky:2016knh} 
proposed to search for evidence for the GW memory effect in a 
population of BBH mergers, for which each individual event is below
the threshold for detection. 
Lasky \textit{et al}.\ showed that for a population of 
GW150914-like events, around 100 BBH mergers are needed to find 
evidence for the GW memory effect in the population.
An important insight in~\cite{Lasky:2016knh} was that only 
a subset of mergers in the population can be used to build evidence
for the GW memory effect, because of degeneracies of certain 
``extrinsic'' parameters (parameters that are \textit{not} the 
masses or spins of the black holes) in the detectors' responses to 
the GWs.
Moreover, a criteria (which can be computed from the GWs) 
was found in~\cite{Lasky:2016knh} to determine whether a 
given detection would be likely to contribute evidence for the GW 
memory in the population or not.

In this paper, we revisit the forecasts in~\cite{Lasky:2016knh}
in light of the nine additional BBH detections after the GW150914
event.
The first ten detections have now allowed models of the distribution
of BBH masses to be constrained by observational 
data~\cite{abbott2019binary}.
We use populations of BBHs consistent with these models to 
estimate the amount of time the advanced LIGO and Virgo detectors
will need to detect the GW memory effect in these 
populations.\footnote{As this work was coming to completion, 
related forecasts for the number of events needed to detect the 
GW memory effect were made in~\cite{Hubner:2019sly}.
We discuss the relationship between~\cite{Hubner:2019sly} and this
paper in greater detail in Sec.~\ref{sec:discussion}.}
We find that, on average, after a five-year observation period, 
the signal-to-noise ratio for the GW memory effect in the population of BBHs will be about three (near the threshold to be observed).
There have been a number of different approximations used to 
compute the GW memory effect from BBH mergers 
(see, e.g.,~\cite{Johnson:2018xly,Favata:2009ii,Favata:2010zu,Talbot:2018sgr}). 
We caution that these models can differ in their 
predictions for the amplitude of the memory effect, and this does
have an impact on the signal-to-noise for the memory effect in
our populations.

The remainder of this article is structured as follows: 
In Sec.~\ref{sec:gwwaveform}, we describe how we calculate the 
relevant gravitational waveforms used throughout this paper.
Sec.~\ref{sec:methods} describes our data-analysis procedures.
Section~\ref{sec:results} contains the main results of our study:
the criteria to determine when a BBH merger will contribute
to building evidence for the memory in the population, and
the forecasts for the time to detection for the GW memory effect
in our simulated populations of BBHs.
We conclude in Sec.~\ref{sec:discussion}.
A few additional results are given in Appendices~\ref{sec:appquadkludge},~\ref{sec:appmemstack}, and~\ref{sec:appmemsign}.
In the remainder of this article geometric units $G = c = 1$ 
are used. 
We use the Planck 2015~\cite{Ade:2015xua} cosmology to associate
a luminosity distance of a BBH to its redshift.

\section{Gravitational Waveform Models}
\label{sec:gwwaveform}

In this section, we discuss several different aspects of the
gravitational waveform models we use throughout this paper:
(i) the conventions for the multipolar expansion of the GW
polarizations, (ii) the specific waveform approximants we use 
in this paper, (iii) the procedure used to calculate the 
waveform associated with the nonlinear GW memory effect, 
(iv) the effect of using different waveform 
approximants in the procedure of (iii), and 
(v) a degeneracy among certain extrinsic 
parameters in the waveform.

\subsection{Gravitational waveforms and their spin-weighted spherical-harmonic expansion}

Gravitational-wave detectors, such as LIGO and Virgo, are not
equally sensitive to the two polarizations of the GWs, which
come from different locations on the sky. 
The sensitivity of the detector to the plus and cross polarizations 
of the gravitational waveform (denoted by $h_{+}$ and $h_{\times}$,
respectively) is given by two antenna response functions, 
$F_{+}(\alpha,\delta,\psi)$ and $F_\times(\alpha,\delta,\psi)$,
which we parameterize by the right ascension $\alpha$, the
declination $\delta$, and the polarisation angle $\psi$. 
[We use the conventions that $\alpha\in(0,2\pi)$, 
$\delta\in(-\pi/2,\pi/2)$, and $\psi\in(0,\pi)$.]
The time-dependent strain measured by the detector is expressed
as the combination of the two polarizations:
\begin{equation}
\label{eq:totstrain}
    h(t) = F_{+}(\alpha,\delta,\psi)h_{+}(t) + F_{\times}(\alpha,\delta,\psi)h_{\times}(t) 
\end{equation}
(see, e.g.,~Appendix~B of~\cite{Anderson:2000yy}). The expressions for $F_+$ and $F_\times$ are taken from the LIGO Algorithm Library (LAL)~\cite{lalsuite}.

For nonprecessing BBH systems, it is convenient to decompose
the complex strain, $h_+ - i h_\times$, using a basis of
spin-weighted spherical harmonics (with spin weight $s=-2$)
that is adapted to the binary.
We assume the binary is in the $x$-$y$ plane, so that the orbital 
(and total) angular momentum points along the $z$ axis.
We denote the spin-weighted spherical harmonics by
${}^{(-2)}Y_{\ell m}(\iota,\phi_c)$.
We choose our coordinates such that $\iota$ represents the angle 
between angular momentum and the line of sight to the detector, and 
$\phi_c$ is the angle between the $x$ axis and the line of site
to the detector projected into the plane of the binary.
The conventions for the harmonics we use are those implemented 
in the \texttt{gwsurrogate} package~\cite{gwsurrogate} (which are 
computed from recurrence relations given in Appendix~B 
of~\cite{Lewis:2001hp}).
The expansion is then given by
\begin{equation}
\begin{split}
\label{eq:hlm}
h_{+} - i h_{\times} = \sum\limits_{l=2}^{\infty}\sum\limits_{m=-\ell}^{m=\ell} &h_{lm}(t;
\vec\sigma) \ ^{(-2)}Y_{\ell m} (\iota,\phi_c) \, ,
\end{split}
\end{equation}
where we have written the spherical-harmonic modes $h_{\ell m}$ of 
the gravitational waveform as a function of time $t$ and a set of
parameters $\vec \sigma$.
For nonprecessing binaries, the parameters included in $\vec\sigma$
are the heavier BH mass $m_1$, the lighter BH mass $m_2$, the 
dimensionless BH spins $\chi_{1z}$ and $\chi_{2z}$ (which are 
assumed to be aligned or anti-aligned with the orbital angular
momentum), and luminosity distance $d_L$. 

This spherical-harmonic decomposition is useful, because for 
nonprecessing BBHs, the amplitudes of the different $(\ell,m)$ 
modes fall off rapidly with $\ell$ and the corresponding frequency 
of the mode is proportional to $m$ (see, e.g.,~\cite{Berti:2007fi}). 
Modes with $m\neq 0$ are referred to as ``oscillatory'' modes.
The dominant oscillatory mode is the quadrupole mode $h_{22}$, 
whereas the modes with $(\ell,|m|) \neq (2,2)$ are notably smaller
in their amplitudes, and are sometimes referred to as ``subdominant''
or ``higher-order'' modes.   
As we discuss in more detail later in this section, for the
quasicircular, aligned-spin binaries considered in this paper, 
the GW memory effect that is computed using $h_{22}$ can be
expanded in just the two modes $h_{20}$ and $h_{40}$ (and the GW
polarization associated with these modes is just the plus polarization)~\cite{Favata:2010zu}.
Thus, modes with $m=0$ are sometimes called ``memory'' 
modes.\footnote{Note, however, that this classification is based 
on the behavior of the waveform modes during the inspiral; during 
the merger and ringdown the $m=0$ modes can have an oscillatory
part, and the memory can appear in modes with $m\neq 0$.}
By parity, it can be shown that for the aligned-spin
binaries considered in this paper the modes with $m<0$ are related 
to the modes with $m>0$ by the relation 
$h_{\ell,m}^* = (-1)^m h_{\ell,-m}$; thus, we will subsequently
only refer to the modes with $m>0$ and not their counterparts with
$m<0$ when discussing which modes we use.

Finally, we conclude this section with a few additional 
definitions that are less standard, but which will be useful
later in this paper.
Let us denote the plus and cross polarizations associated with
a given mode $h_{lm}$ as:
\begin{subequations} \label{eq:hlmPlusCross}
\begin{align} 
   h_+^{\ell m} &:= \Re{h_{\ell m}{^{(-2)}Y_{\ell m}}} \, ,\\
   h_\times^{\ell m} &:= -\Im{h_{\ell m}{^{(-2)}Y_{\ell m}}} \, .
\end{align}
\end{subequations}
Similarly, let us define the strain measured by the detector for 
a particular mode $h_{lm}$ as
\begin{equation} \label{eq:hlmDet}
   h_{(lm)}(t) := F_{+}(\alpha,\delta,\psi) h_+^{lm}(t) + F_{\times}(\alpha,\delta,\psi) h_\times^{lm}(t) \, .
\end{equation}
Thus the full GW strain measured by the detector can be
written as
\begin{equation}
    h(t) = \sum_{\ell,m} h_{(lm)}(t) \, .
\end{equation}
While the quantities $h_{(lm)}(t)$ are not something that would be
easily measurable by GW detectors like LIGO and Virgo for a single
$(\ell,m)$ mode, they will be useful for explaining certain
degeneracies that occur when the GWs measured by a GW detector
are influenced predominantly by a few individual $h_{(lm)}(t)$
in the total strain $h(t)$.

\subsection{Computing the oscillatory waveform modes}
\label{sec:surmodel}

To compute the dominant and higher-order oscillatory waveform modes, 
we use the NRHybSur3dq8 surrogate model~\cite{Varma:2018mmi}.
This model can be used to generate waveforms from BBHs with 
mass ratios $q$ in the range $q = m_1/m_2\leq 8$ and with aligned
spins with magnitudes $|\chi_{1z}|,|\chi_{2z}| \leq 0.8$.
The model was built from a catalog of spinning, non-precessing numerical relativity (NR)
simulations~\cite{Boyle:2019kee} that were 
``hybridized''~\cite{Ajith:2007qp} with post-Newtonian
(PN) (see e.g. the review article \cite{Blanchet:2013haa} and references therein) and effective-one-body (EOB) waveforms~\cite{Buonanno:1998gg, Bohe:2016gbl}. 
The surrogate model is a type of interpolant (based on reduced-order
modeling techniques \cite{Field:2013cfa,Cannon:2012gq,Purrer:2014fza,Blackman:2015pia, Blackman:2017dfb})
that allows the waveform model to be rapidly evaluated with high
accuracy in its range of validity.

We use the Python package \texttt{gwsurrogate}~\cite{gwsurrogate} to evaluate
the NRHybSur3dq8 surrogate model.
This model includes $(\ell,m)$ modes with $2\leq\ell\leq 4$ 
[though not the (4,0) or (4,1) modes] and the (5,5) mode.
We restrict to generating the dominant mode $h_{22}$ and the five higher-order modes $h_{21}, h_{32}, h_{33}, h_{44}$ and $h_{55}$.
We neglect the other modes, as they are either small or 
not well resolved in the NR simulations. 
We choose the duration of the waveform to be such that the 
$h_{55}$ mode starts at a frequency of $f_0 = 10$ Hz, for
all the binaries (of different masses) that we consider. 

\subsection{Computing the nonlinear GW memory}
\label{sec:memcalc}

The GW memory effect can be computed from NR simulations using
the technique of Cauchy-characteristic extraction
(see, e.g.,~\cite{Bishop:1996gt}) as was done in~\cite{Pollney:2010hs}
for a few nonprecessing, equal-mass BBHs. 
The more commonly used methods of waveform extraction (and 
extrapolation), however, fail to resolve the effect (see, 
e.g.,~\cite{Boyle:2019kee}).
The memory effect is required by the conservation of supermomentum 
(the conserved quantity associated with the supertranslation 
symmetries of the Bondi-Metzner-Sachs group); thus, the memory can 
be computed approximately from the gravitational waveform model
without the GW memory effect by determining the waveform required
to maintain supermomentum conservation (see, 
e.g.,~\cite{Nichols:2017rqr,Ashtekar:2019viz,Compere:2019gft}).

While supermomentum conservation provides the theoretical 
underpinning for the approximate method for computing the GW
memory effect from waveforms without GW memory, the resulting
prescription can be described in simpler terms: 
One can compute the nonlinear GW memory following the same
procedure used to calculate linear memory from massless fields
after replacing the material stress-energy tensor with the 
effective stress-energy tensor of gravitational
waves~\cite{Thorne:1992sdb}.
The derivation of the result relies on solving the relaxed Einstein
equations (in harmonic gauge), and has been given in several 
places (e.g.,~\cite{Wiseman:1991ss,Favata:2008yd}); as a result,
we do not rederive the result, but quote the final result 
instead.

The strain associated with the memory effect can be computed from
the expression
\begin{equation}
\label{eq:memcor}
    h_{jk}^\textrm{TT,mem} = \frac{4}{r} \int\limits_{-\infty}^{u} du' \left[  \int \frac{d E}{d\Omega' du} \frac{n'_jn'_k}{1- \textbf{n}' \cdot \textbf{n}} d\Omega' \right]^{TT} \, .
\end{equation}
In this expression, we have defined the retarded time $u$, the
distance to the source $r$, the unit vector pointing from the
source $\mathbf n = \mathbf{x}/r$, the 
solid-angle element $d\Omega$, and the GW luminosity per solid 
angle $dE/(du d\Omega)$.
To relate the expression in Eq.~\eqref{eq:memcor} to the two
polarizations of the GWs, it is necessary to contract 
Eq.~\eqref{eq:memcor} with the complex polarization tensor
as follows:
\begin{equation} \label{eq:hmemPols}
    h_+^\mathrm{mem} - i h_\times^\mathrm{mem} = 
    h_{jk}^\textrm{TT,mem} (e_+^{jk} - i e_\times^{jk}) \, .
\end{equation}
It is convenient to define the polarization tensors using a 
complex vector $m^*_j$ (where $*$ denotes complex conjugation).
In spherical coordinates, $(\iota,\phi_c)$, the vector $m^*_j$ 
is given by 
$m^{*j} = [(\partial_\iota)^j - i \sin\iota (\partial_{\phi_c})^j]/\sqrt 2$, and the polarization tensor 
is then
\begin{equation}
e_+^{jk} - i e_\times^{jk} = m^{*j} m^{*k} \, .
\label{eq:eplus_cross}
\end{equation}

For practical computations of the nonlinear memory effect, it is
common to expand the energy flux in terms of the time-derivatives
of the GW strain expanded in spin-weighted spherical harmonics:
\begin{equation}
\label{eq:fluxharmonic}
    \frac{d E}{du d\Omega} = \frac{r^2}{16\pi}\sum\limits_{\ell',\ell'',m',m''} \Big \langle \dot{h}_{\ell' m'}\dot{h}^{*}_{\ell''m''}  \Big \rangle \ {^{(-2)}Y_{\ell' m'}}\  ^{(-2)}Y_{\ell''m''}^{*}.
\end{equation}
The angle brackets around the term 
$\dot{h}_{\ell' m'}\dot{h}^{*}_{\ell''m''}$ mean to average over 
a few wavelengths of the radiation.
By substituting Eq.~\eqref{eq:fluxharmonic} into 
Eq.~\eqref{eq:memcor}, the memory waveform becomes a sum over
products of two spin-weighted spherical harmonics.
However, it is then useful to expand 
$h_+^\mathrm{mem} - i h_\times^\mathrm{mem}$ in spin-weighted 
spherical harmonics as
\begin{equation} \label{eq:hmempx}
    h_+^\mathrm{mem} - i h_\times^\mathrm{mem} = 
    \sum_{\ell,m} h_{\ell m}^\mathrm{mem} \ ^{(-2)}Y_{\ell' m'} \, ,
\end{equation}
so that the multipole moments $h_{\ell m}^\mathrm{mem}$ 
are functions of time that are determined by a double angular
integral of products of three spin-weighted spherical harmonics.
These integrals, although somewhat complicated, can be evaluated
numerically (as was done in~\cite{Talbot:2018sgr}).
Alternately, the integral can be recast in terms of
symmetric-trace-free tensors or scalar spherical harmonics 
and evaluated analytically (in terms
of Clebsch-Gordon coefficients or Wigner 3-j 
symbols)~\cite{Favata:2008yd,Faye:2014fra,Nichols:2017rqr}.

We compute several appoximate expressions for the polarizations in
Eq.~\eqref{eq:hmempx} in the next part of this section.

\subsection{Memory waveform models}
\label{sec:memmodels}

We describe in this part three different approximate methods that
have been used to compute the GW memory effect from BBHs.
Two of the models differ only in the number of spherical-harmonic
modes included in the expansion of the GW luminosity in 
Eq.~\eqref{eq:fluxharmonic}.
The other model uses additional approximations that we
will discuss in greater detail herein. 
We discuss one additional waveform model in
Appendix~\ref{sec:appquadkludge} that is used to compute the GW 
memory effect in~\cite{Johnson:2018xly}.
This model makes several additional approximations, which have
the effects of decreasing the amplitude of the GW memory effect
by a factor of roughly two, and introducing a small oscillatory
part that would not be expected in these particular 
spherical-harmonic modes of the memory effect.
For these reasons, we do not include this model in the calculations
in this part.
In Fig.~\ref{fig:mem_models}, we provide an example that shows 
that the three different approximations can lead to results that 
differ by several tens of a percent.

\subsubsection{Descriptions of waveform models}

\paragraph{Quadrupole approximation} 
In~\cite{Favata:2009ii}, Favata considered the memory generated
by the luminosity just from the spherical-harmonic mode 
$h_{22}$ in the luminosity in Eq.~\eqref{eq:fluxharmonic}. 
However, unlike previous work in the PN approximation 
(e.g.,~\cite{Wiseman:1991ss,Kennefick:1994nw}), 
Ref.~\cite{Favata:2009ii} used the full inspiral-merger-ringdown
waveform for $h_{22}$, which was fit to results from a NR simulation.
The angular integral in Eq.~\eqref{eq:memcor} can be performed
straightforwardly in this approximation, and the GW memory 
is predominantly in the spherical-harmonic
modes $h_{20}^\mathrm{mem}$ and $h_{40}^\mathrm{mem}$ (and is
thus plus polarized).\footnote{Note, that there can also be small
contributions to the mode $h_{44}$, though this will be ignored in
this approximation.}
The resulting waveform for the plus polarization of the 
GW memory effect can be written as
\begin{equation}
\label{eq:mem}
    h^\mathrm{mem}_+(u) = \frac{r}{192 \pi}\sin^2\iota(17+\cos^2\iota)\int\limits_{-\infty}^{u} |\dot{h}_{22}|^2 du' \, .
\end{equation}
We evaluate the mode $h_{22}$ using the surrogate model
described in Sec.~\ref{sec:surmodel}.
We will call the waveform computed via this procedure the
``quadrupole'' model. 

\paragraph{Minimal waveform model}
Favata also constructed what he called the ``minimal waveform model''
(MWM) in~\cite{Favata:2009ii}.
The MWM is an analytical approximation to the time-domain quadrupole 
GW memory waveform, which is based on using the PN approximation
to the waveform during the inspiral and a superposition of 
quasinormal modes during the merger and ringdown.
It was then calibrated (by a constant rescaling) to match with 
the memory computed from an early effective-one-body (EOB) 
waveform model tuned to NR simulations \cite{Damour:2008te}. 
The time-domain MWM also has a Fourier transform that can be 
computed analytically to give an analytic frequency-domain 
waveform for the GW memory effect.
This allows the MWM to be computed rapidly, which has made it
useful in studies that perform Bayesian inference using 
GW memory models (e.g.~\cite{Lasky:2016knh,Yang:2018ceq}).
However, the EOB model~\cite{Damour:2008te} against which the MWM 
was calibrated was not as precisely tuned to NR, and it 
overestimates the amplitude of the GW memory effect (this was
noted in~\cite{Favata:2009ii} and also in \cite{Talbot:2018sgr}). 
Nevertheless, because the MWM is a common approximation, we
include it as the second of our three approximate methods.

\paragraph{Quadrupole and higher multipole model}
The waveform from the GW memory effect had been computed to 3PN 
order in~\cite{Favata:2008yd}, and at this order in the PN 
approximation, subdominant modes of the oscillatory GW strain 
enter into Eq.~\eqref{eq:fluxharmonic}.
The PN approximation only holds during the inspiral of a BBH,
so it was only more recently in Talbot 
\textit{et al.}~\cite{Talbot:2018sgr} that higher multipole moments
were included for the full inspiral-merger-ringdown waveforms 
used to calculate the GW memory.
Specifically, in~\cite{Talbot:2018sgr}, higher-order GW modes up 
to (and including) $\ell=4$ 
were used in Eq.~\eqref{eq:fluxharmonic} to compute 
$h_+^\mathrm{mem} - i h_\times^\mathrm{mem}$. 
Including the higher-order modes resulted in a roughly ten-percent
increase in the amplitude of the GW memory effect, for comparable
mass binaries. 
Accompanying the paper~\cite{Talbot:2018sgr} was the release of
a Python package called \texttt{gwmemory}~\cite{gwmemory}. 
We compute the GW memory waveform using this package with the 
oscillatory GW modes generated by the surrogate waveform model 
in Sec.~\ref{sec:surmodel} (we leave out the $h_{55}$ mode, 
because the \texttt{gwmemory} package does not 
compute the angular integrals for oscillatory modes with
$l\geq 5$). 
We will refer to this third memory waveform model as the 
``higher-mode'' model.

\subsubsection{Illustration of differences between waveform models}

\begin{figure}
  \centering
  \includegraphics[width=0.99\columnwidth]{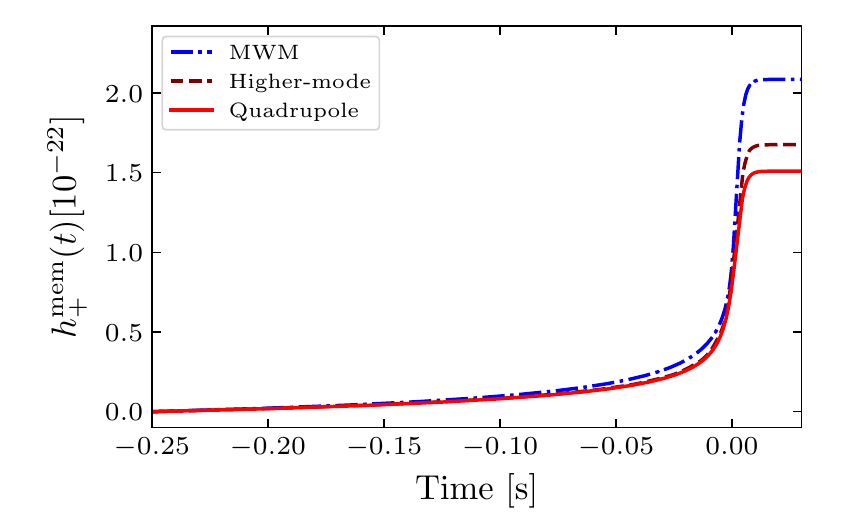} 
  \includegraphics[width=0.99\columnwidth]{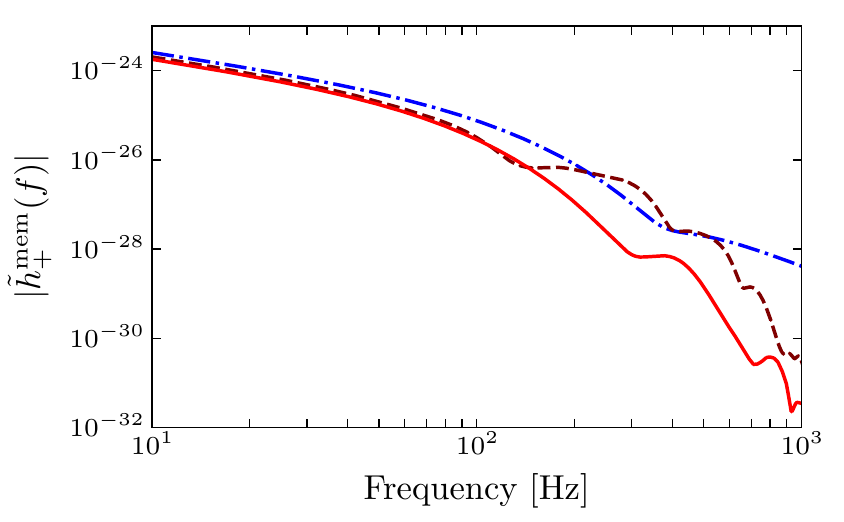}
    \caption{
    Gravitational waveforms associated with the nonlinear GW memory
    effect for a BBH with masses $m_1 = 30 M_{\odot}$ and 
    $m_2 = 30 M_{\odot}$, at a luminosity distance 
    $d_L = 500 \mathrm{Mpc}$ and at an inclination $\iota = \pi/2$.
    The three curves are three different approximations for 
    computing the GW memory waveform: the blue dotted-dashed line 
    is the MWM of~\cite{Favata:2009ii}, the red solid line is the 
    quadrupole approximation also in~\cite{Favata:2010zu}, and the 
    dashed brown line is the higher-mode model
    of~\cite{Talbot:2018sgr} (see the the text for more detailed
    descriptions of the models).
    \textit{Top}: The time-domain waveform $h^{\mathrm{mem}}_{+}(t)$ 
    for the nonlinear GW memory for the three models.
    \textit{Bottom}: The nonlinear GW memory waveform in the 
    frequency domain for the three models.} 
    \label{fig:mem_models}
\end{figure}

We now show the differences that arise from using the
different prescriptions for the three gravitational waveform
models of the GW memory effect for a typical stellar-mass 
BBH system.
Figure~\ref{fig:mem_models} shows $h_+^\textrm{mem}(t)$ (top panel) 
and $|\tilde h_+^\mathrm{mem}(f)|$ (bottom panel) for the three 
GW-memory waveform models for a BBH with masses 
$m_1 = 30 M_{\odot}$ and $m_2 = 30 M_{\odot}$.
We choose the luminosity distance to be $d_L = 500 \mathrm{Mpc}$ and 
the inclination angle to be $\iota = \pi/2$ (we replace $r$ with
$d_L$ for binaries at cosmological distances).
We set the value of $h_+^\mathrm{mem}(t)$ to be zero at the 
starting time in the top panel of Fig.~\ref{fig:mem_models},
for this comparison.~\footnote{The MWM includes an initial offset 
from zero that is computed from the PN approximation, while the
other models do not.}
We compute the Fourier transform $\tilde h_+^\mathrm{mem}(f)$
from $h_+^\mathrm{mem}(t)$ in the following ways:
For the MWM, we use the analytical expression given 
in~\cite{Favata:2009ii}; for the other two models, we pad
the time domain waveform, window the time-domain waveform 
with a Planck window~\cite{McKechan:2010kp} to remove edge effects, 
and use the fast Fourier transform (FFT) 
algorithm~\cite{Cooley:1965zz} implemented in 
\texttt{NumPy}~\cite{numpy,van_der_Walt_2011}.

While the time dependence of the three models is similar,
the amplitudes are not.
The quadrupole and higher-mode models are similar (they differ
in the constant value of $h_+^\mathrm{mem}(t)$ at late times $t$ 
by around ten percent).
These two models, however, differ from the MWM by a larger amount.
This difference is also present in the frequency domain waveforms,
although it is more difficult to observe in the bottom panel
of Fig.~\ref{fig:mem_models}.

The higher-mode model of~\cite{Talbot:2018sgr} is expected to 
be the most accurate of the three, because it introduces the 
fewest assumptions and approximations.
However, it is also the slowest to compute, because it involves
the largest number of waveform modes.
Because the quadrupole approximation of~\cite{Favata:2009ii}
differs by a relatively small amount and is faster to compute, 
we will use this waveform for most of our forecasts in 
Sec.~\ref{sec:results}; however, this will slightly underestimate
the signal to noise of the memory effect in the population of BBHs.
The MWM would typically overestimate it instead (we describe 
this in more detail in Sec.~\ref{sec:gw150914}).

\subsection{Degeneracies between waveform parameters}
\label{sec:wavedeg}

We discuss in this section properties of the quantities
$h_{(lm)}(t)$ introduced in Eq.~\eqref{eq:hlmDet} that will
affect whether a given detection will be likely to contribute
any significant evidence for the GW memory in the population
of binaries (similarly to what was done in~\cite{Lasky:2016knh}).

In GW parameter estimation, it is well known that there are 
strong correlations between some parameters measured from a
BBH merger by interferometric detectors when performing parameter
estimation using just the dominant $l=2$, $m=2$ waveform mode
(e.g., the correlation between inclination $\iota$ and luminosity
distance $d_L$~\cite{Cutler:1994ys}).
It is also well known, however, that by including higher-order modes 
in the waveform model, some of these correlations can be broken 
and improved constraints on the parameters of the gravitational waveform model can be obtained~\cite{Graff:2015bba, Payne:2019wmy, London:2017bcn, Kumar:2018hml, Chatziioannou:2019dsz}. 

One salient type of correlation for detecting the GW memory 
effect was noted by Lasky \textit{et al.} in~\cite{Lasky:2016knh}:
namely, they described a degeneracy for the dominant $l=2$, $m=2$
under transformations of the form 
\begin{equation} \label{eq:PsiPhic}
    (\psi,\phi_c)\rightarrow (\psi',\phi_c') = 
    (\psi+\pi/2,\phi_c+\pi/2) \, .
\end{equation}
The quantity $h_{(22)}(t)$ was invariant, but other modes
$h_{(lm)}(t)$ were not.
The reason for the degeneracy of $h_{(22)}(t)$ is straightforward
to understand: At fixed sky location $(\alpha,\delta)$, the
antenna patterns $F_+$ and $F_\times$ are periodic in the
polarization angle $\psi\in(0,\pi)$; thus, the transformation
$\psi\rightarrow\psi+\pi/2$ changes the sign of the antenna
patterns $F_+\rightarrow -F_+$ and $F_\times\rightarrow -F_\times$.
Because the polarizations associated with the mode $h_{22}$
satisfy $h^{22}_+ -i h^{22}_\times \propto e^{2i\phi_c}$, then
under the transformation $\phi_c\rightarrow\phi_c+\pi/2$ it follows
that $h^{22}_+ -i h^{22}_\times \rightarrow 
-(h^{22}_+ -i h^{22}_\times)$.
This leaves the mode $h_{(22)}(t)$ invariant under this 
transformation.

For the purposes of discussing some of the correlations we have
found in this work, it will be useful to consider the slightly
more general transformation
\begin{equation} \label{eq:FpxPhic}
    (F_+,F_\times,\phi_c)\rightarrow (\pm F_+,\pm F_\times,\phi_c+\beta)
    \, .
\end{equation}
For a general mode $h_{(\ell m)}(t)$, a straightforward 
calculation then shows that under the 
transformation~\eqref{eq:FpxPhic}, the mode transforms as
\begin{equation} \label{eq:hlmFpxPhic}
\begin{split}
    h_{(\ell m)}(t) \rightarrow & \pm
    [F_+h^{\ell m}_\times(t) - F_\times h^{\ell m}_+(t)]\sin m\beta
    \\
    & \pm h_{(\ell m)}(t) \cos m\beta \, .
\end{split}
\end{equation}
The case $m=2$, $\beta=\pi/2$, and the sign flip for
$F_+$ and $F_{\times}$
recovers the degeneracy of the mode $h_{(22)}(t)$ 
discussed in detail above [and the expression
above shows it is actually valid for any mode with $m=2$ (mod 4)
and $\beta=\pi/2$].

We conclude this part by noting a few other degeneracies, and
how these degeneracies can be broken.
For example, the degeneracy of the mode $h_{(22)}(t)$ discussed
above is not shared by a number of other modes.
For example, when $m=1$ or $m=3$ (mod 4) and $\beta=\pi/2$,
then the $h_{(\ell m)}(t)$ transform in a nontrivial
way under~\eqref{eq:FpxPhic}.
Similarly, when $m=0$ (mod 4) and $\beta=\pi/2$, then the expression 
in Eq.~\eqref{eq:hlmFpxPhic} reduces to
$h_{(\ell m)}(t) \rightarrow \pm h_{(\ell m)}(t)$.
Thus, the presence of any of these higher-order modes 
$h_{(\ell m)}(t)$ in $h(t)$ (including the 
memory modes with $m=0$) will break this degeneracy in 
Eq.~\eqref{eq:PsiPhic}.
If $h(t)$ has as its dominant two modes $h_{(22)}(t)$ 
and $h_{(44)}(t)$, then the degeneracy in 
Eq.~\eqref{eq:PsiPhic} will still be broken; however, there will 
be an additional degeneracy
under the transformation that leaves $F_+$ and $F_\times$
invariant and has $\beta=\pi$ (this is relevant for 
Fig.~\ref{fig:mem_overlap_3030}).
Finally, it need not be simply the transformation
$\psi \rightarrow \psi + \pi/2$ that changes the sign of
the antenna patterns $F_+$ and $F_\times$.
In Appendix~\ref{sec:appmemsign} is an example of
an unfortuitous sky location that leads to an additional 
degeneracy among the right ascension and declination.

\section{Methods for assessing the presence of the GW memory effect in a BBH population}
\label{sec:methods}

In this section, we discuss how we assess when the GW memory 
effect is present in a population of BBHs.
Specifically, we describe computing signal-to-noise ratios (SNRs)
for individual events and for populations of events, and determining 
a criteria like that used in~\cite{Lasky:2016knh} for when a 
given event will contribute significant SNR towards finding the 
memory effect in the population of BBHs.
In connection with this last point, we discuss Bayesian inference.

\subsection{Computing signal-to-noise ratios}

In the context of GW data analysis, matched filtering is 
an important component of finding and assessing the significance 
of a GW signal that is buried in what is generally assumed to be 
stationary Gaussian noise of a GW detector 
(see, e.g.,~\cite{Finn:1992wt,Cutler:1994ys}).
The filter involves cross correlating the detector output
$d(t)=s(t) + n(t)$ [where $s(t)$ is the GW signal and $n(t)$ is 
the detector noise] with a bank of template gravitational 
waveforms, $h(t)$.
This cross correlation of the data and a template is most
conveniently written in terms of the noise-weighted inner product 
in the frequency domain as
\begin{equation}
\label{eq:overlap}
\langle d|h \rangle = 4\Re \int\limits_{f_0}^{f_1} df  \frac{\tilde{d}^*(f)\tilde{h}(f)}{S_n(f)},
\end{equation}
Here $\tilde{d}(f)$ and $\tilde{h}(f)$ are the Fourier transforms 
of the $d(t)$ and $h(t)$, respectively, and $S_n(f)$ is the 
one-sided power spectral density (PSD) of the detector's noise. 
The frequencies $f_0$ and $f_1$ define, respectively, the lower 
and upper range of the detector's sensitivity in the frequency
domain.
The inner-product of the signal with itself is the square of the
optimal signal-to-noise ratio (SNR):
\begin{equation}
\label{eq:snr_opt}
\rho^2 = \langle s|s \rangle \, .
\end{equation}
Equations~\eqref{eq:overlap} and~\eqref{eq:snr_opt} are important
elements in assessing the likelihood of a signal existing in 
noisy data (see, e.g.,~\cite{Finn:1992wt} for more detail).

The signal $s(t)$ is generally not known \textit{a priori},
so in practice, the square of the SNR is estimated from the
data by computing the inner product of the data $d(t)$
with a family of templates $h(t;\vec{\theta})$ with different 
parameters $\vec{\theta}$. 
If Bayesian parameter estimation is used to construct a
posterior probability distribution $p(\theta|d)$ for the parameters
$\vec\theta$, then in general, there will be a distribution of 
squared matched filter SNRs, which can be computed from
\begin{equation}
\label{eq:SNRdistribution}
    \rho^2(\vec\theta;d) = \langle d|h(\vec\theta) \rangle\, ,
\end{equation}
where $h(\vec\theta)$ are the template waveforms that are 
consistent with the parameters $\vec\theta$ of the posterior
probability distribution. 
For a single noise realization, the median of this distribution 
will not necessarily equal the optimal SNR given by 
$\sqrt{\langle s|s \rangle}$.
However, for an event with a high SNR in Gaussian noise, 
the expected value of this median will be the
optimal SNR, when averaging over different noise realizations.
This will not necessarily be the case for signals with low 
SNRs or signals that transform nontrivially under some of the
degeneracies discussed in Sec.~\ref{sec:wavedeg} (as we will
discuss in more detail in Sec.~\ref{sec:signmeasurement}).

In this paper, we will compute a number of SNRs.
For individual events in a single GW detector these SNRs are
as follows: (i) the SNR $\rho_i^\mathrm{osc}$, which corresponds
to the SNR for the $i^\mathrm{th}$ GW detector (LIGO Hanford, 
LIGO Livingston, or Virgo) from the oscillatory GW modes 
for a BBH described in Sec.~\ref{sec:surmodel}; 
(ii) the SNR $\rho_i^\mathrm{22}$, which is the SNR of just 
the dominant $\ell=2$, $m=2$ mode; 
(iii) the SNR $\rho_i^\mathrm{hom}$, which contains all the
oscillatory modes except for $h_{22}$ and $h_{32}$; and
(iv) the SNR $\rho^\mathrm{mem}_i$, which is the SNR for 
the GW memory signals described in Sec.~\ref{sec:memmodels}.
For signals that are measured in $N_d$ detectors (where $N_d = 3$ 
in this paper), the network SNR is typically taken to be the
sum in quadrature of the individual detector SNRs for each 
relevant class of signal:
\begin{equation}
\label{eq:SNRnetwork}
 \rho^2_N =  \sum_{\text{i}=1}^{{N_d}}\rho^2_\text{i} \, .    
\end{equation}
This is a reasonable definition when the Gaussian noise in each
detector is independent of the other detectors.

As a practical matter, we compute the SNRs for all the different
types of GW signals as follows.
If the signal is not already in the frequency domain, we take 
the time-domain signal and apply a Planck
window~\cite{McKechan:2010kp} before computing the Fourier 
transform.
The upper limit of the relevant integral is taken to be the 
Nyquist frequency, (i.e., half of our sampling frequency 
$f_s = 8192$Hz).
The lower limit is chosen to be $f_0 = 10$ Hz, which
is the low-frequency cutoff of the noise curves that
we use for the PSDs of the Advanced LIGO and Virgo detectors.
Specifically, we use the PSDs for aLIGO and Virgo at their
design sensitivities, which are given 
in~\cite{TheLIGOScientific:2014jea, TheVirgo:2014hva}.

\subsection{The case for combining subthreshold 
GW-memory-effect signals}
\label{sec:stacking}

For non-precessing BBH mergers, there is a clear hierarchy of 
SNRs for the dominant, higher-order, and memory modes
of the waveform (i.e., $\rho_N^\mathrm{22} > 
\rho_N^\mathrm{hom} > \rho_N^\mathrm{mem}$).\footnote{For example,
for an event consistent with 
GW150914~\cite{abbott2016observation,TheLIGOScientific:2016wfe,ligo2018gwtc}, the SNRs are
$\rho_N^\mathrm{osc}=76.5$, $\rho_N^\mathrm{22}=75.9$, 
$\rho_N^\mathrm{hom}=3.9$, and $\rho_N^\mathrm{mem}=0.21$ 
(for the quadrupole memory waveform model).}
As a rough rule of thumb, when the SNR of a particular signal 
is less than around one, that signal is sufficiently weak that neither 
can it be claimed to be detected, nor can much be inferred about
it from the data.
More concretely, if, for example, $\rho_N^\mathrm{osc}$ exceeds
the threshold of detection, but $\rho_N^\mathrm{mem}$ is less
than one, then neither would it be possible to claim detection for
the memory effect, nor would including the memory effect in 
waveforms used for parameter estimation inform the posterior
distributions for the parameters in the waveform model.
At the same time, when there is a confident detection of the dominant
mode $h_{22}$, but the SNR for the higher-order modes does not itself
pass the SNR threshold, the higher-order modes may still influence
the estimation of parameters if $\rho_N^\mathrm{hom}$ is still
greater than around one.
This last point has important implications for detecting the GW
memory effect, which were noted in~\cite{Lasky:2016knh}, because
the higher-order modes turn out to be useful for breaking some
of the degeneracies mentioned in Sec.~\ref{sec:wavedeg}.

Given what is currently known about the population of BBHs 
from the LIGO and Virgo observations~\cite{abbott2019binary},
the sensitivities of the Advanced LIGO and Virgo detectors,
and the relative strengths of the dominant, higher-order, and
memory modes, it is expected that the detected events will fall
into the following classes of SNRs for the different modes:
(i) a significant fraction of the events will have 
$\rho_N^\mathrm{22}$ passing the threshold for detection, $\rho_N^\mathrm{hom} < 2$, and $\rho_N^\mathrm{mem} < 1$;
(ii) a smaller fraction of events will have $\rho_N^\mathrm{22}$ 
passing the threshold for detection, $\rho_N^\mathrm{hom} > 2$ but
under the threshold of detection, and $\rho_N^\mathrm{mem} < 1$; and
(iii) a handful of events with $\rho_N^\mathrm{22}$ passing the threshold for detection, $\rho_N^\mathrm{hom} > 2$, and $\rho_N^\mathrm{mem} > 1$ (some of these events may have 
$\rho_N^\mathrm{hom}$ near the threshold for detection, but
it is not expected that $\rho_N^\mathrm{mem}$ will reach this
level).
More detailed numbers for specific BBH populations are given in
Sec.~\ref{sec:bbhpop}.

Because there are expected to be a large number of detections of
BBHs, and because the SNR of the memory effect is not expected to
exceed the threshold for detection, it seems reasonable to follow
the approach of Lasky \textit{et al.}~\cite{Lasky:2016knh}, who
proposed combining multiple BBH detections with subthreshold memory
signals to provide evidence for the effect in a population of 
BBH mergers. 
To assess whether the memory was present in the 
population,~\cite{Lasky:2016knh} use two methods, one based on
computing evidence ratios for signals with and without memory modes,
and a second based on computing the SNR of the memory in the 
population of BBH events.
We more closely follow the second approach of~\cite{Lasky:2016knh}
based on the total SNR of $N_e$ events measured in network of
$N_d$ detectors.
(though we briefly discuss the relationship between the two 
methods in Sec.~\ref{sec:discussion}).

Assuming that all the events are independent, the noise in the
network is Gaussian, and the signal in the data is known exactly,
then the total SNR for the memory waveforms in the population of 
BBHs is given by 
\begin{equation}
 \rho_{\text{tot}} = \sqrt{\sum_{\text{j}=1}^{N_e}(\rho^{\mathrm{mem}}_{N,j})^2} 
 \, .
 \label{eq:memsum_ideal}
\end{equation}
Here $\rho^{\textrm{mem}}_{N,j}$ is the network SNR of the 
GW memory effect in the detector network for the j$^\mathrm{th}$ 
detection.
The SNR in Eq.~\eqref{eq:memsum_ideal} will grow approximately with 
the number of detections and detectors as 
$\sqrt{N_e N_d}$.\footnote{This growth is also explained in
Appendix~\ref{sec:appmemstack}, using an analogy based on stacking 
GW memory waveform signals.}
If the memory signals for each event is known, then the
quantity $\rho^{\textrm{mem}}_{N,j}$ could be computed
from Eqs.~\eqref{eq:snr_opt} and~\eqref{eq:SNRnetwork}.
However, when the signal is not known \textit{a priori}, one might
instead consider using the median value of the SNR in
Eqs.~\eqref{eq:SNRdistribution} to determine the network 
SNR~\eqref{eq:SNRnetwork} of the memory effect
for the j$^\mathrm{th}$ detection.

Using the median value in Eq.~\eqref{eq:SNRdistribution} leads to
certain complications, because of the hierarchy of SNRs 
described in this part and the degeneracies among the
waveform parameters discussed in Sec.~\ref{sec:wavedeg}.
Consider, for example, the case when 
$\rho_N^\mathrm{22}$ passes the threshold for detection, $\rho_N^\mathrm{hom} < 2$, and $\rho_N^\mathrm{mem} < 1$.
Because of the degeneracies described in Sec.~\ref{sec:wavedeg},
then the SNRs for the parameters $\vec\theta$ and 
$\vec\theta'$ [defined by the transformation in
Eq.~\eqref{eq:PsiPhic}] will satisfy
$\rho^2_\mathrm{mem}(\vec\theta';d) \approx 
-\rho^2_\mathrm{mem}(\vec\theta;d)$.
Thus, both sets of parameters will be nearly equally consistent 
with the observed data, and the distribution of matched-filter
SNRs for the GW memory effect will contain significant support
for both positive and negative values (so that the median
would be close to zero).
However, as Lasky \textit{et al.} observed in~\cite{Lasky:2016knh},
there could be a sufficient number of events with 
$\rho_N^\mathrm{22}$ passing the threshold for detection,
$\rho_N^\mathrm{hom} > 2$ but under the threshold of detection, 
and $\rho_N^\mathrm{mem} < 1$.
For these events, the higher-order modes can break the degeneracies
that allow strains $h_\textrm{mem}$ with opposite signs to be 
consistent with the data, so that the true value of the SNR
will be close to the median value (this will also be shown in more
detail in Sec.~\ref{sec:signmeasurement}).

Therefore, if one were to use all detected events, $N_e$, to
estimate the total SNR for the memory effect in the population,
$\rho_\mathrm{tot}$, in Eq.~\eqref{eq:memsum_ideal}, this 
would generally overestimate the SNR, because of the degeracies
discussed in the previous paragraph.
Instead, there are two approximations that one could
make to obtain a more realistic estimate of the SNR 
$\rho_\mathrm{tot}$:
The first would be to replace $\rho^{\mathrm{mem}}_{N,j}$
in Eq.~\eqref{eq:memsum_ideal} with the median value that is 
consistent with the posterior distribution of parameters of the
waveform.
We will \textit{not} take this approach in this paper.
Rather, we will instead follow a procedure like that 
in~\cite{Lasky:2016knh}, in which we will only consider 
those events in which $\rho_N^\mathrm{hom}$ satisfies a
SNR threshold cut (similarly to in~\cite{Lasky:2016knh}, 
we choose this to be $\rho_N^\mathrm{hom} > 2$, for reasons
which we discuss more in Sec.~\ref{sec:signmeasurement}).
Thus, we will estimate the SNR of the memory effect in a
BBH population by
\begin{equation}
 \rho_{\text{tot}} = \sqrt{\sum_{\text{j}=1}^{N'_e}(\rho^{\textrm{mem}}_{N,j})^2}
 \, ,
 \label{eq:memsum}
\end{equation}
where $N'_e$ is the number of detected events that satisfy
our SNR cut for the higher-order modes, and
$(\rho^{\textrm{mem}}_{N,j})^2$
is computed from Eqs.~\eqref{eq:snr_opt} and~\eqref{eq:SNRnetwork}.

Note that choosing the hard cut of $\rho_N^\mathrm{hom} > 2$
may underestimate the total SNR of the memory effect, because some
events near the threshold, but that do not make the cut could still
contribute a reduced, though nonzero, SNR.
Thus, we will typically compute the SNR using both 
Eqs.~\eqref{eq:memsum_ideal} and~\eqref{eq:memsum} as ways of 
roughly estimating a lower and upper bound for the memory SNR.
In the next part of this section, we discuss our choice for the SNR
threshold in more detail, and the Bayesian methods that we
used to determine this criteria.

\subsection{Determining the ``sign'' of the memory effect through inference of the source parameters}
\label{sec:PE}

In~\cite{Lasky:2016knh}, Lasky \textit{et al.} determined a
criteria based on the SNR required in a particular combination
of higher-order modes with $m=1$ and $m=3$ which broke the
degeneracy in Eq.~\eqref{eq:PsiPhic} for the mode $h_{(22)}$.
In terms of the somewhat more general transformation in
Eq.~\eqref{eq:FpxPhic}, the degeneracy we would like to break
is that between the two signs of the antenna patterns 
$\pm F_+$ and $\pm F_\times$ for some of the specific 
angles $\beta$ discussed in Sec.~\ref{sec:wavedeg}.
The reason for this is as follows: 
In the quadrupole approximation the strain $h_+^\textrm{mem}$
[computed from Eq.~\eqref{eq:mem}] is non-negative and is
independent of $\beta$; thus, the sign of memory strain measured 
by a GW detector, $h_\textrm{mem}$, is completely determined by 
the sign of the antenna pattern $F_+$.
We will sometimes then refer to the breaking of the
degeneracy in Eq.~\eqref{eq:FpxPhic} [or~\eqref{eq:PsiPhic}]
as determining the ``sign'' of the memory effect 
(or just ``the memory sign''), as was done in~\cite{Lasky:2016knh}.
Rather than using the specific combination of higher-order 
modes used in~\cite{Lasky:2016knh}, we will base our criteria
on the SNR $\rho_N^\mathrm{hom}$ [as we discussed in 
Sec.~\ref{sec:wavedeg}, other modes besides the $m=1$ and
$m=3$ modes can break the degeneracy in Eq.~\eqref{eq:FpxPhic}
for particular values of $\beta$].

To determine whether we can measure the sign of the memory effect,
therefore, we need to determine how accurately we can measure the
four parameters that determine the degeneracy 
in Eq.~\eqref{eq:FpxPhic}: right ascension $\alpha$, declination
$\delta$, polarization $\psi$, and phase $\phi_c$.
For most sufficiently well localized sources at most sky locations,
this degeneracy reduces to resolving the degeneracy between 
the latter two parameters $\psi$ and $\phi_c$. 

To ascertain how well we can recover the unknown signal parameters,
we use Bayesian inference~\cite{Bayes:1764vd, JaynesProbabilityTheory, Veitch:2014wba} to compute
posterior probability density functions (PDFs) for the relevant
parameters. 
Specifically, given the detector output $d$ and a signal hypothesis $H$ that involves a set of parameters $\vec{\theta}$, we compute
the posterior PDF for the parameters $\vec\theta$ via 
Bayes' theorem:
\begin{equation} \label{eq:Bayes}
    p(\vec{\theta}|d,H) \propto \mathcal{L}(d|\vec{\theta},H)p(\vec{\theta}|H) \, .
\end{equation}
Here $p(\vec{\theta}|H)$ is the prior PDF for the parameters $\vec{\theta}$ and 
$\mathcal{L}(d|\vec{\theta},H) = p(d|\vec{\theta},H)$ is 
the likelihood function. 
For our detector network, we assume that the noise is Gaussian
and that the noise in each detector is uncorrelated with the other 
detectors.
This implies we can write the joint likelihood as the product of 
the individual likelihoods~\cite{Cutler:1994ys,Veitch:2014wba}:
\begin{equation}
    \mathcal{L}_N = \prod_{\text{i}=1}^{{N_d}} \mathcal{L}_i(d_i|\vec{\theta},H).
\end{equation}
The log of the individual likelihoods of the data in each detector 
given some signal model $H \equiv h(\vec{\theta})$ is 
given by (see, e.g.~\cite{Cutler:1994ys})
\begin{equation}
\label{eq:loglike}
    \log \mathcal{L} ( d |  \vec{\theta}, H) \propto 
    -\frac{1}{2}\sum_{\text{i}=1}^{{N_d}} \bigg \langle d_i- h_i(\vec{\theta}) \ | \ d_i- h_i(\vec{\theta}) \bigg \rangle \, .
\end{equation}

For a BBH in a quasicircular orbit, there are 15 parameters in
$\vec{\theta}$. 
We will restrict to nonspinning binaries for our parameter
estimation studies, which reduces the dimension of the 
parameter space to nine. 
Because the degeneracy is among the extrinsic parameters and there
generally are not strong correlations between intrinsic and 
extrinsic parameters~\cite{Cutler:1994ys},
we fix the component masses of the binary to their true values. 
This leaves us with the extrinsic parameters, given by the 
set $\vec{\theta} = \{d_L,\iota,\alpha,\delta,\psi,\phi_c,t_c\}$.
The last parameter $t_c$, which had not been introduced previously,
is the time at coalescence. 
For each of these extrinsic parameters we specify the prior PDFs
to be uninformative priors.
Specifically, we take the priors for the source's sky location to 
be isotropic and the luminosity distance to be uniform in volume.
The width of the distance prior is adjusted to cover a sufficiently large range around the true luminosity distance of the binary. 
For the prior on the orientation of the binary with respect to the
line of sight, we again assume an isotropic prior.
Finally, we take the priors for the polarization to be uniform
in $(0,\pi)$ and for the coalescence time $t_c$ to be uniform in
a 200 ms.~window centered on the true value.

Although, in general, the detector output consists of both the
GW strain and a realization of Gaussian noise, for our parameter
estimation studies, we do not include any noise.
The intent of this approximation is to better understand the
correlations and degeneracies among parameters as a function
of $\rho_N^\mathrm{hom}$ without introducing a bias from a 
specific noise realization (though with noise, one may
sometimes require a higher value of $\rho_N^\mathrm{hom}$ 
to break the degeneracies).
The results we find without noise also should be similar to 
those that would be obtained from averaging over many random 
Gaussian noise realizations with zero mean. 
The detector noise is taken into account when calculating the 
noise-weighted inner product in Eq.~\eqref{eq:loglike}, because
it involves the noise power spectral density of the GW detectors.

We use the ensemble MCMC sampler \texttt{kombine}~\cite{kombine} 
to determine the posterior PDF from Eq.~\eqref{eq:Bayes}
for detector data consisting of the waveform from a binary with 
parameters $\vec{\theta}_{\rm true}$.
From the posterior PDF, we can determine if the degeneracy between
$\psi$ and $\phi_c$ is broken (i.e., if the PDFs of these angles 
will be concentrated around the true values) or not (i.e., there is 
similar support in the posterior PDFs for $\psi$ and $\phi_c$ and
both values shifted by $\pi/2$). 
This will determine how conclusively we can measure the sign
of the memory for this particular binary.
With the posterior PDFs, we can then also compute for each point
$\vec{\theta}_s$ in the parameter space the associated GW memory
waveform $h_{\textrm{mem}}(\vec{\theta}_s)$ using the 
quadrupole model, and the corresponding estimate for the square of 
the SNR in Eq.~\eqref{eq:SNRdistribution}.
We will discuss the results of these calculations in the 
next section.

\section{Results: Memory sign and forecasts for detection of the
GW memory effect}
\label{sec:results}

In the first part of this section, we illustrate how the 
presence of higher-order GW modes allows the sign of the
GW memory effect to be measured.
In the next two sections, we highlight the number of detections
and the amount of detector time necessary to detect the
GW memory effect in several different types of populations of
BBHs.
We first consider a BBH population of GW150914-like events
followed by two classes of BBH populations that are consistent 
with the models of the BBH populations computed by the LIGO and
Virgo Collaborations in~\cite{abbott2019binary}.

\subsection{Measuring the memory sign}
\label{sec:signmeasurement}

\begin{figure*}
  \centering
  \subfigure{\includegraphics[width=0.99\columnwidth]{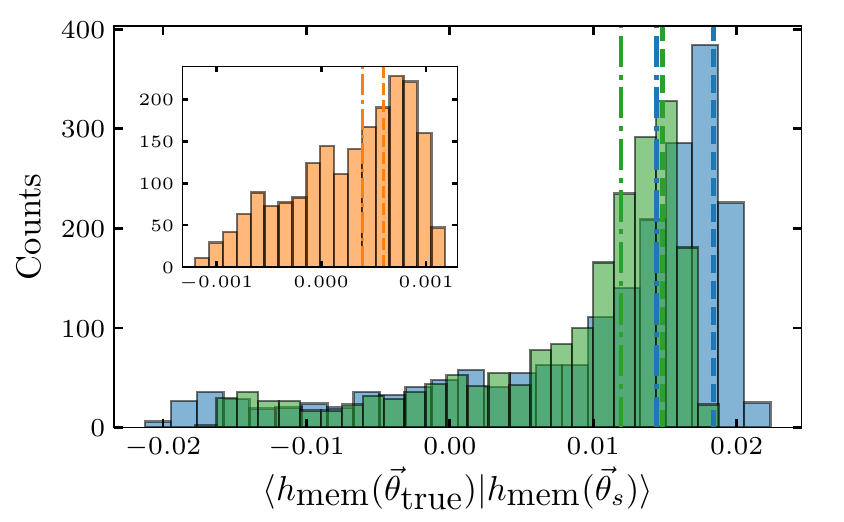}}\hfill
  \subfigure{\includegraphics[width=0.99\columnwidth]{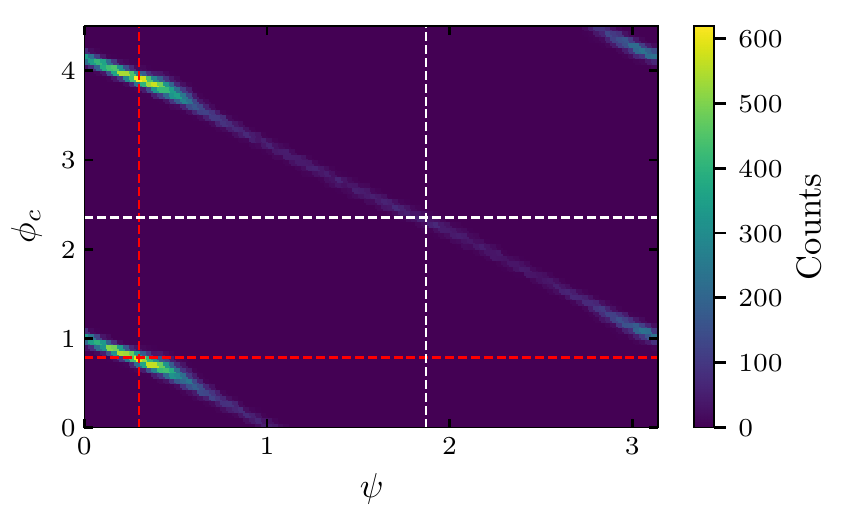}}\hfill
  \subfigure{\includegraphics[width=0.99\columnwidth]{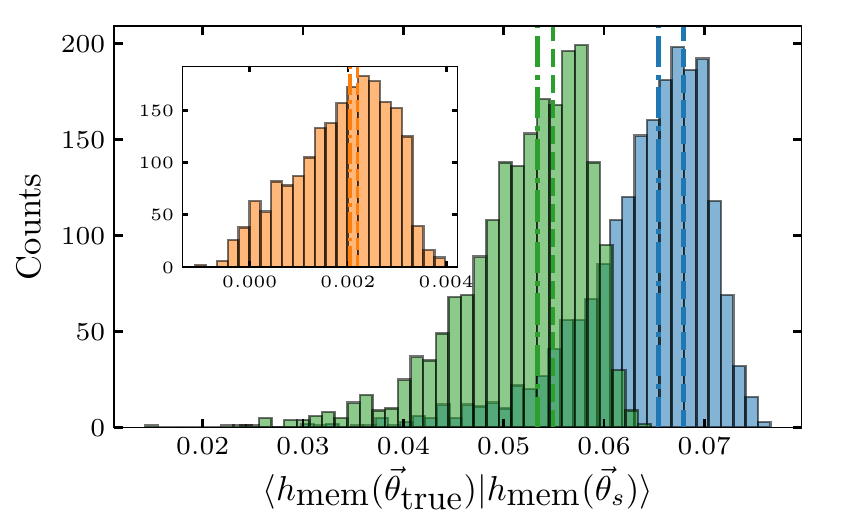}}\hfill
  \subfigure{\includegraphics[width=0.99\columnwidth]{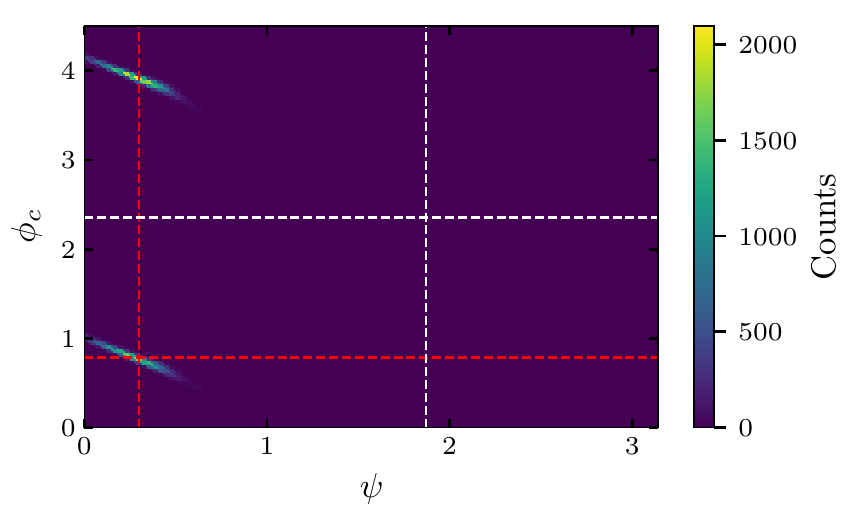}}\hfill
  \subfigure{\includegraphics[width=0.99\columnwidth]{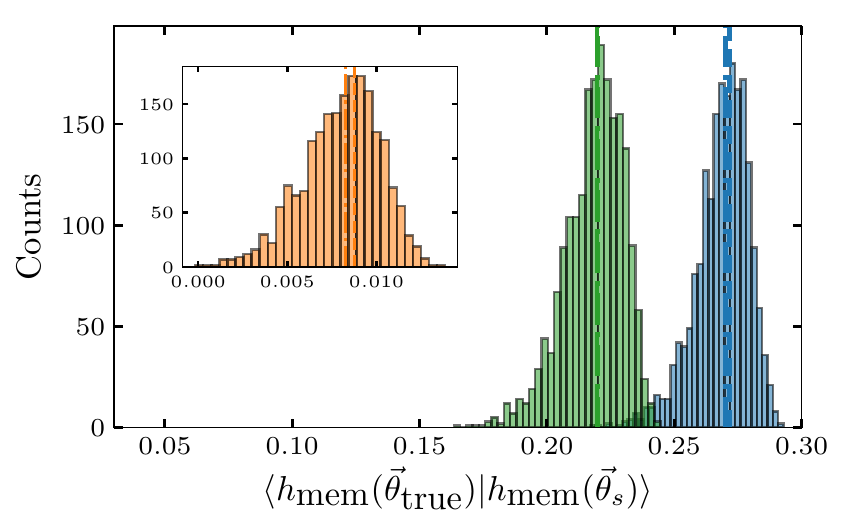}}\hfill
  \subfigure{\includegraphics[width=0.99\columnwidth]{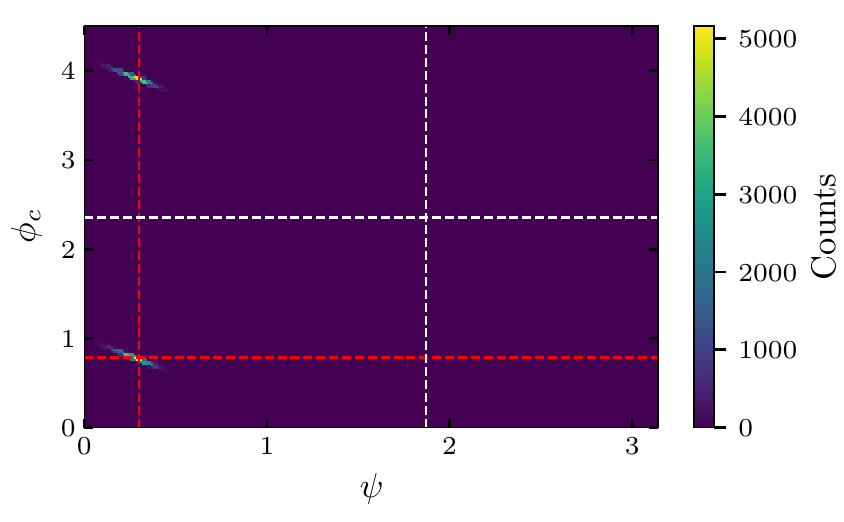}}\hfill
    \caption{Posterior distributions (not normalized) for a nonspinning, 
    equal-mass binary with $m_1=m_2= 30 M_\odot$ with extrinsic
    parameters given by $\alpha = 4.2$ rad, $\delta = -0.8$ rad, $\psi = 0.3$ rad and $\iota = 2.1$ rad.
    \textit{Left column}: The overlap between the true signal, $h_{\textrm{mem}} (\vec{\theta}_{\textrm{true}})$, and
    templates consistent with the posteriors, 
    $h_{\textrm{mem}} (\vec{\theta}_s)$, for three different 
    SNRs in the higher-order modes,
    $\rho_N^{\mathrm{hom}}$.
    From top to bottom the SNRs are 
    $\rho_N^{\mathrm{hom}} \approx 1$, 2, and 4; the different
    SNRs were obtained by varying the luminosity distance of 
    the source, while keeping other parameters fixed. 
    The inner product for advanced LIGO Hanford, advanced LIGO 
    Livingston and Virgo are shown in green, blue, and orange,
    respectively. 
    The vertical dashed lines represent the optimal $\rho^2$ injected values and the vertical dotted-dashed lines represent the median values of the distributions.
    \textit{Right column}: The 2D PDFs for $\psi$ and $\phi_c$, 
    for the same binaries in the corresponding rows. 
    Red dashed lines show the ``true'' injected values and
    the white dashed line shows the degenerate values.
    Already at an SNR of 1, the $\psi$-$\phi_c$ degeneracy is 
    partially broken, whereas for the SNRs 2 and 4 in the middle and
    bottom rows, the degeneracy is broken, and the sign of the
    detector's response to the memory effect is well known for
    all detectors.}
    \label{fig:mem_overlap_3030}
\end{figure*}

In principle, it would be possible to perform Bayesian inference
on every binary in a population of BBHs, to determine whether
we can confidently determine the median network SNR of each GW 
memory effect in the data.
Because of the significant computational cost of doing this, 
similarly to in~\cite{Lasky:2016knh}, we instead look for a 
criteria based on $\rho_N^\mathrm{hom}$ that will be 
satisfied when we confidently know the sign of the memory, which
we can then use in lieu of full Bayesian parameter estimation. 
To establish this criterion, we perform Bayesian inference on a 
handful of BBHs of different masses, sky locations, polarizations,
and orientations of the binaries.
A representative result for an equal mass BBH is shown 
in Fig.~\ref{fig:mem_overlap_3030}.
We use this result to demonstrate that the criteria of
$\rho_N^\mathrm{hom} > 2$ will be sufficient for most BBHs.
However, we caution that there can be small regions of the 
extrinsic parameter space where this criteria is not as 
strong, for particular sky locations.
One such example is shown for a BBH with $q = 3/2$  
in Appendix~\ref{sec:appmemsign}. 

For each binary, we tune the amplitude of $\rho^{\textrm{hom}}_N$ 
either by changing the luminosity distance $d_L$ or the 
inclination $\iota$ (in the former approach, all the SNRs of 
the different modes scale inversely with the distance in the
same way, whereas in the latter approach, the relative amplitudes
of the SNRs of the different modes change much more).
We run Bayesian parameter estimation, as described in 
Sec.~\ref{sec:PE}, to determine how large $\rho^{\textrm{hom}}_N$
must be to break the degeneracies and to determine the sign of
the GW memory effect. 

We show the results for an equal mass BBH with 
$m_1 = m_2 = 30 M_{\odot}$ in Fig.~\ref{fig:mem_overlap_3030}. 
The three rows correspond to three luminosity distances 
$d_L = 1250$, 650, and 325 Mpc (going from top to bottom);
the corresponding SNRs $\rho^{\textrm{hom}}_N$ are given by
$\rho^{\textrm{hom}}_N \approx 1$, 2 and 4, respectively. 
This binary is detectable by the advanced detector network 
at all three distances, because the oscillatory SNRs are
roughly 25, 48 and 96. 
The left column shows the inner product of the data [the
true signal $h_{\textrm{mem}} (\vec{\theta}_{\mathrm{true}})$]
with templates that are consistent with the posteriors,
$h_{\textrm{mem}} (\vec{\theta}_{\mathrm{s}})$.
The blue and green histograms correspond to this inner product for
the advanced LIGO Livingston and Hanford, respectively, and the 
inset orange histogram shows this for the Virgo detector.
The right column shows the 2D posterior PDF for the parameters
$\psi$ and $\phi_c$.

When $\rho^{\textrm{hom}}_N \approx 1$, we find that the degeneracy 
in Eq.~\eqref{eq:PsiPhic} between $\psi$ and $\phi_c$ is not 
fully broken; thus, there is nontrivial support for both signs
of noise-weighted inner product 
$\langle h_{\mathrm{mem}} (\vec{\theta}_{\mathrm{true}}) |
h_{\mathrm{mem}} (\vec{\theta}_{\mathrm{s}}) \rangle$ 
in the left panel. 
This occurs because although the true values of $\psi$ and $\phi_c$
are favored (indicated by the intersection of the red dashed lines),
there is also some support for the true values both shifted by
$\pi/2$ (indicated by the intersection of the white dashed lines). 
The presence of the negative noise-weighted inner product is most
obvious for Virgo (in the inset), where the amplitude of the inner
product is smallest; however, it is also visible in the histograms
for LIGO-Hanford and LIGO-Livingston, despite the larger amplitude
for the inner product.

For $\rho^{\textrm{hom}}_N \approx 2$, almost all templates $h_{\textrm{mem}} (\vec{\theta}_s)$ consistent with the 
posterior PDFs have the correct sign, which occurs because the
degeneracy of Eq.~\eqref{eq:PsiPhic} is now almost fully broken.
For $\rho^{\textrm{hom}}_N \approx 4$, $\psi$ and $\phi_c$ are even
better constrained, and the overlap for all detectors is closely 
centered around the optimal SNR squared.
Note that there is a remaining degeneracy between 
$\phi_c$ and $\phi_c + \pi$ apparent in the 2D posteriors 
even at the large values of $\rho^{\textrm{hom}}_N$.
This occurs because the majority of the SNR in 
$\rho^{\textrm{hom}}_N$ comes from $h_{(44)}$ (this was noted
in Sec.~\ref{sec:wavedeg}).
This residual degeneracy does \textit{not} affect the sign of the 
GW memory effect, however.

The results in Fig.~\ref{fig:mem_overlap_3030} are representative
of the required network SNR in the higher-order modes, 
$\rho^{\textrm{hom}}_N$, that is needed to break the degeneracies
that determine the sign of the memory in at least one detector
(though see Appendix~\ref{sec:appmemsign} for an example of a 
very specific sky location and polarization that requires a 
slightly higher value of $\rho^{\textrm{hom}}_N$).
Thus, we conclude that binaries for which the network SNR 
$\rho^{\textrm{hom}}_N \geq 2$ is sufficient to be able to 
determine the memory sign.
As a result, we will use this criteria to determine when we 
include a given detection in the total SNR for the memory in
Eq.~\eqref{eq:memsum} in a BBH population.
This criteria is used throughout the next two subsections.

\subsection{GW150914-like binary-black-hole population}
\label{sec:gw150914}

Before we investigate different populations from those studied
in~\cite{Lasky:2016knh}, we first aim to understand the effects
of using a different waveform model and a slightly different 
criteria for the SNR in the higher-order GW modes on the same
population of BBHs used by~\cite{Lasky:2016knh}.
Specifically, we consider in this section a population of
GW150914-like binaries.
These are nonspinning binaries with $m_1 = 36 M_{\odot}$, 
$m_2 = 29 M_{\odot}$ and $d_L = 410$ Mpc, which are values
consistent with GW150914~\cite{TheLIGOScientific:2016wfe}. 
The rest of the binary's parameters are distributed uniformly
in $\alpha$, $\sin \delta$, $\cos \iota$, $\psi$ and $\phi_c$. 
In this analysis, as in Ref. \cite{Lasky:2016knh}, we use a
detector network of the two LIGO detectors at design
sensitivity~\cite{TheLIGOScientific:2014jea}, and we use a 
network SNR for the oscillatory part of the signal of 12 as our
threshold for detection (i.e., $\rho^{\textrm{osc}}_N \geq 12$). 

We calculate the GW memory waveforms for all detections using the
three different waveform models described in Sec.~\ref{sec:memmodels}.
For each model, we calculate the associated total memory SNR 
from Eq.~\eqref{eq:memsum_ideal} for a population of 
100 GW150914-like binaries. 
For detections with $\rho_N^{\textrm{hom}} \geq 2$ (where just
higher-order modes with $l \leq 3$ and odd $|m|$ are used), 
we include the network SNR for the memory effect in the sum, and
for the remaining detections, we set
$\rho^{\mathrm{mem}}_\mathrm{N,j} = 0$,
for each of the waveform models (as was described in 
Sec.~\ref{sec:signmeasurement}).\footnote{Although we do not
include GW modes $l > 3$ in the oscillatory waveform to match 
with~\cite{Lasky:2016knh}, the higher-mode memory waveform is
calculated using all modes up to $l=4$ as stated in 
Sec.~\ref{sec:memmodels}.}

\begin{figure}[htb]
    \includegraphics[width=0.99\columnwidth]{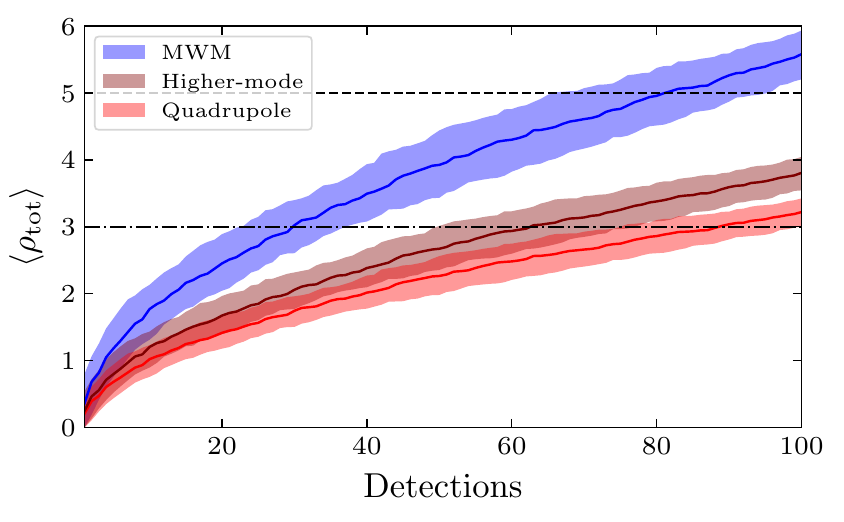}
    \caption{The total memory SNR versus the detection number for 
    a population of GW150914-like binaries computed with the three
    GW memory waveform models in Sec.~\ref{sec:memmodels}. 
    The solid lines are the median values over 100 realizations
    of this population, and the shaded regions are 1-$\sigma$
    confidence intervals.
    The colors red, maroon, and blue correspond to the quadrupole,
    higher-mode, and MWM models, respectively. 
    The dotted-dashed and dashed black lines show two SNR 
    thresholds used in~\cite{Lasky:2016knh}.
    Our calculations with the MWM are consistent with the ones
    in ~\cite{Lasky:2016knh} (which also used the MWM).
    The other two models have notably smaller SNRs
    for the memory effect.}
    \label{fig:gw150914-growth}
\end{figure}

We repeat the above analysis for 100 realizations of this 
GW150914-like population (and we use the same realizations
for the three different waveform models).
Figure~\ref{fig:gw150914-growth} shows how the total SNR for
the memory effect grows over the 100 detections. 
The solid lines show the median SNR over the 100 realizations of the
population, $\langle \rho_{\text{tot}}\rangle$, and the shaded
regions indicate the 1-$\sigma$ confidence intervals (i.e., the
symmetric, 68\% credible region). 
The three colors (blue, maroon, and red) correspond to the
three different waveform models described in 
Sec.~\ref{sec:memmodels} (the MWM, higher-mode, and quadrupole, respectively).

\begin{table}[htb]
    \centering
    \caption{Total memory SNR $\langle \rho_{\text{tot}}\rangle$ for 
    the three different waveform models of Sec.~\ref{sec:memmodels}
    after 30 and 90 detections.
    The numbers are the median value, and the error bars are
    1-$\sigma$ confidence intervals.}
    \begin{ruledtabular}
    \begin{tabular}{ccccc}
       Detection Number  & MWM & Higher-mode & Quadrupole 
       \\
       \hline
       30 & $3.10_{-0.41}^{+0.33}$ & $2.11_{-0.28}^{+0.23}$ & $1.79_{-0.23}^{+0.19}$   \\
       90 & $5.30_{-0.36}^{+0.36}$ & $3.61_{-0.25}^{+0.24}$ &
       $3.06_{-0.21}^{+0.21}$  \\
    \end{tabular}
    \end{ruledtabular}
    \label{tab:GW150914}
\end{table}

For reference, we give the median value of the 
SNR $\langle \rho_{\text{tot}}\rangle$ and the 1-$\sigma$ 
confidence intervals for the population after 30 and 90 
detections in Table~\ref{tab:GW150914}.
We choose these numbers, because they are round numbers 
where the median value of $\rho_{\text{tot}}$ for the MWM 
(the model used in~\cite{Lasky:2016knh}) is close to the 
two values of $3$ and $5$ used for the thresholds of 
detection in~\cite{Lasky:2016knh} (which are intended to 
represent 3- and 5-$\sigma$ significant detections of the memory
effect in the GW150914-like population).
Our results for the MWM are similar to those found 
in~\cite{Lasky:2016knh}.
As Table~\ref{tab:GW150914} shows, the higher-mode and quadrupole
models (which make fewer assumptions when computing the
GW memory effect) produce significantly smaller values for
the total memory SNR in the GW150914-like population.
We consider the results of these models to be more representative
of the GW memory signal (for the reasons discussed in 
Sec.~\ref{sec:memmodels}), so we expect the total SNR of the
memory in this population of BBHs to be closer to these values.

We conclude this part by noting that for this GW150914-like
population, on average two-thirds of the detections pass the SNR
threshold in the higher-order modes.

\subsection{Power-law mass-function populations}
\label{sec:bbhpop}

Nine additional BBHs were detected by the LIGO-Virgo
Collaboration after GW150914, in the first two observing runs, and 
these nine detections informed models of the population of 
BBHs~\cite{abbott2019binary}.
We now repeat our calculations of 
$\langle \rho_\mathrm{tot} \rangle$ for populations that are
consistent with the models in~\cite{abbott2019binary}. 

\subsubsection{Simulated BBH populations}

Specifically, we use model A of~\cite{abbott2019binary} for 
the distribution of the BH masses in a BBH system. 
For this model, the mass ratio of the binary $q$ is assumed to 
follow a uniform distribution; the distribution of the primary
component mass, $m_1$, is taken to be a power-law (with index
$\alpha_{\rm pow}$) and the mass range is restricted between
$m_{\text{min}}$ and $m_{\rm max}$.
This means that the mass distribution can be written in the form
\begin{equation}
\label{eq:powerlaw}
    p(m_1,m_2 \ | m_{\rm min}, m_{\rm max}, \alpha_{\rm pow}) =  C(m_1) m_1^{-\alpha_{\rm pow}},
\end{equation}
for $m_{\text{min}} \leq m_{2} \leq m_{1} \leq m_{\text{max}}$ 
[where $C(m_1)$ is the normalization] and the probability is
zero outside this mass range. 
The minimum black hole mass $m_{\text{min}}$ is fixed to be 
$5.0 M_{\odot}$, so there are two free parameters in the mass
distribution: $m_\mathrm{max}$ and $\alpha_\mathrm{pow}$.
The parameters $\alpha_{pow}$ and $m_{\text{max}}$ were inferred 
in~\cite{abbott2019binary} by assuming that the GW detections in 
the first and second observing runs followed a Poisson process
with an unknown rate per comoving volume of BBH mergers, $R$.
The three parameters were jointly inferred from the GW detections
using Bayesian techniques.
The median values of the mass-distribution parameters are 
$\alpha_\mathrm{pow} = 0.4$ and $m_{\text{max}} = 41.6 M_{\odot}$, 
while the rate per volume's median value is
$R = 64.9 \ \mathrm{Gpc}^{-3} \mathrm{yr}^{-1}$.
These three parameters are correlated in nontrivial ways; 
see~\cite{abbott2019binary} for more detail. 

We also allow the BHs in our population to have aligned spins.
We again use the results of~\cite{abbott2019binary} to 
determine the distribution of spin parameters.
Specifically, we assume that the aligned-spin magnitudes of each BH 
in the binary are independent of one another, and we assume that 
they follow the nonparametric binned distribution illustrated in
the bottom panel of Fig.~7 of~\cite{abbott2019binary}.
This model favors small aligned spins, so we do not expect the
results to differ much from a population of nonspinning BBHs.

Because the surrogate model is valid for a subset of the
allowed mass ratios and spins, we restrict to aligned-spin binaries
with mass ratios $1 \leq q \leq 8$ and dimensionless spin magnitudes
$|\chi_{1z}|,|\chi_{1z}| \leq 0.8$. 
We generate BBH mergers uniformly in comoving volume up to 
$d_L = 2\mathrm{Gpc}$ (we do not observe a significant change 
in the total memory SNR by increasing $d_L$).
The remaining extrinsic parameters of the binary are distributed 
in the same way as they were for the BBH population in 
Sec.~\ref{sec:gw150914}. 

Because of the large range of masses and distances for the 
binaries in this BBH population, a more significant number
of the simulated BBHs will not reach the SNR threshold
for detection.
We select the criteria for detection as follows:
For the detector network, we choose the two Advanced LIGO 
detectors and the Virgo detector at their design
sensitivities~\cite{TheLIGOScientific:2014jea, TheVirgo:2014hva}. 
We consider a BBH merger to be detected if the three-detector 
network SNR satisfies $\rho^{\textrm{osc}}_N \geq 8$ and if 
the single-detector SNRs satisfy $\rho^{\textrm{osc}} \geq 4.5$ 
for LIGO and $\rho^{\textrm{osc}} \geq 3.0$ for Virgo. 
To determine binaries for which we know the memory effect's sign, 
we use the criteria $\rho_{N}^{\textrm{hom}} \geq 2$, as in
Sec.~\ref{sec:gw150914}, although we now use all modes mentioned in 
Sec.~\ref{sec:surmodel} except for the $l = 3, m = 2$ mode which 
does not break the degeneracies mentioned in Sec.~\ref{sec:wavedeg}.

The LIGO and Virgo detector network is not operational 
for all times, but just for a fraction of the time (which gets
called the network's ``duty cycle'').
We therefore keep only the fraction of the detections consistent
with the duty cycle of the three-detector network, which based 
on~\cite{detectorstatus} is 50\% (i.e., we exclude 50\% of the
binaries that make the SNR cut for detection).
We calculate the total memory SNR for the population in two 
ways.
As a more conservative estimate, we use Eq.~\eqref{eq:memsum} 
to compute the SNR from the binaries that pass all three
detection, duty-cycle, and higher-mode-SNR cuts.
As an upper bound, we also calculate the total memory SNR in the 
same way except that we include the binaries that do \textit{not} pass the higher-mode-SNR cut 
(this was also done in~\cite{Lasky:2016knh}). 
To determine the uncertainty arising from different realizations
of the population, we generate 300 realizations, and we compute
the median values and confidence intervals for the SNR
over these different realizations. 
We use the quadrupole memory waveform model to model the memory
effect in these simulated populations. 

We simulate each realization for an observation period of five 
years; with the assumed duty cycle, this corresponds to 2.5 years 
of coincident data for the three-detector network.
We find it more useful to compute the total SNR
of the memory effect as a function of observation time, because
both the number of detections and the types of detected binaries
will vary over different realizations of the population, even
for fixed values of the parameters $\alpha_\mathrm{pow}$, 
$m_{\mathrm{max}}$, and $R$.

We perform two types of analyses with the BBH population 
based on model A, which differ only in how we treat the
parameters $\alpha_\mathrm{pow}$, $m_{\mathrm{max}}$, and $R$.
First, we fix the parameters $\alpha_\mathrm{pow}$, 
$m_{\mathrm{max}}$, and $R$ to their median values and sample 
the masses from 300 different realizations of populations with 
these median parameters.
This highlights the uncertainty from different realizations of
a fixed population.
However, there are also uncertainties on 
the merger rate, the maximum mass, and the power law index.
Thus, for our second analysis, we let the values of 
$\alpha_{\rm pow}, m_{\text{max}}$ and $R$ be drawn randomly 
from their respective posterior distributions given 
in~\cite{abbott2019binary}.
This allows us to understand how the total GW memory SNR varies because 
of the uncertainty in the three parameters $\alpha_\mathrm{pow}$,
$m_{\mathrm{max}}$, and $R$.

\subsubsection{SNR for the GW memory effect}

\begin{figure}[htb]
\centering
  \subfigure{\includegraphics[width=0.99\columnwidth]{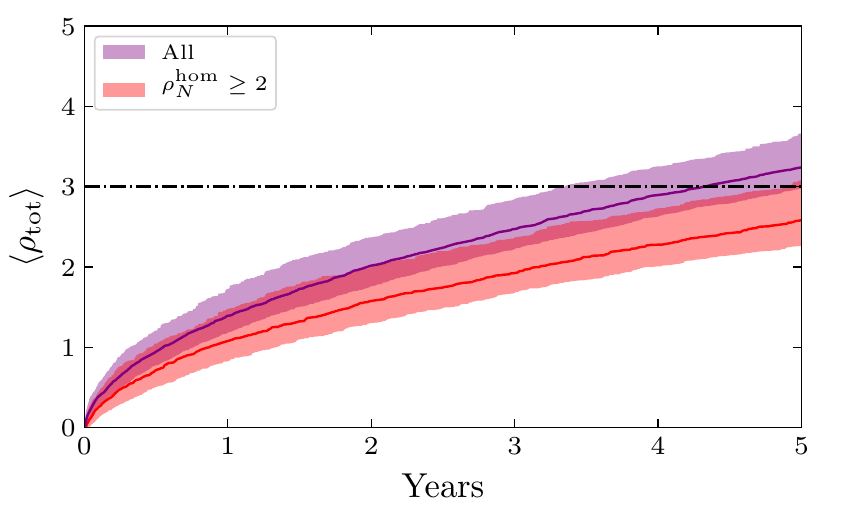}}
  \subfigure{\includegraphics[width=0.99\columnwidth]{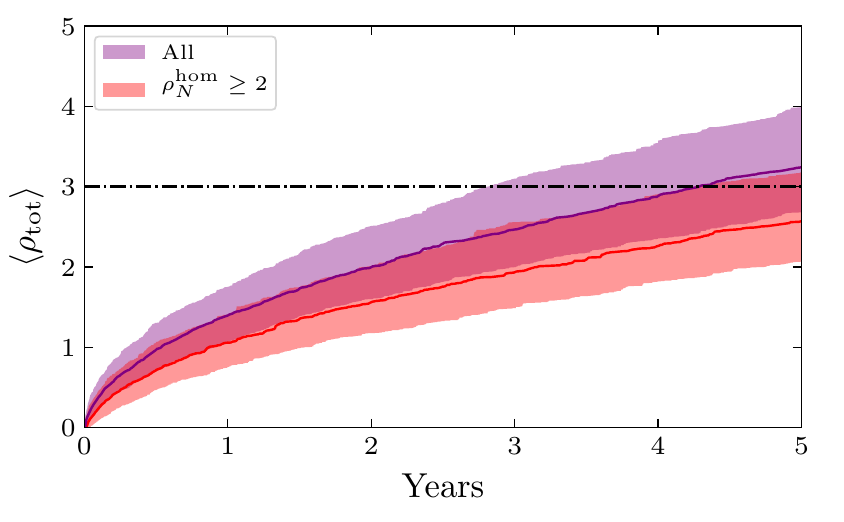}}
    \caption{The total memory SNR for two populations of BBHs 
    generated using model A of~\cite{abbott2019binary} 
    (see the text of Sec.~\ref{sec:bbhpop} for further details). 
    The memory SNR is calculated using the quadrupole memory model. The shaded regions indicate the 
    1-$\sigma$ confidence intervals and the solid lines show the 
    median SNR $\langle \rho_{\text{tot}}\rangle$. In red, only the detections are included which pass our higher-mode-SNR cut $\rho^{\rm hom}_N \geq 2$. In purple, all detections are included.  
    \textit{Top}: The population parameters $R$, $\alpha_{\rm pow}$, 
    and $m_{\rm max}$ are fixed to their median values. 
    \textit{Bottom}: $R$, $\alpha_{\rm pow}$ and $m_{\rm max}$ are drawn according to the distributions given in~\cite{abbott2019binary}. In both analyses, top and bottom panel, the GW memory effect is on the verge of being detected after five years of operation of the advanced LIGO and Virgo detector network at design sensitivity.}
\label{fig:popa-growth}
\end{figure}

\begin{table}[htb]
    \centering
    \caption{Total memory SNR $\langle \rho_{\text{tot}}\rangle$ for 
    the two different analyses of Sec.~\ref{sec:bbhpop}
    after 5 years of detector operation time.
    The numbers are the median value, and the error bars are
    1-$\sigma$ confidence intervals.}
    \begin{ruledtabular}
    \begin{tabular}{ccc}
       Population A  & $\rho^{\rm hom}_N \geq 2$ & All 
       \\
       \hline
       Fixed & $2.58_{-0.32}^{+0.50}$ & $3.24_{-0.26}^{+0.42}$   \\
       Varied & $2.57_{-0.50}^{+0.61}$ & $3.24_{-0.56}^{+0.75}$ \\
    \end{tabular}
    \end{ruledtabular}
    \label{tab:PopA}
\end{table}

Figure~\ref{fig:popa-growth} shows the total memory SNR gained versus
detector operation time in years. 
The shaded regions indicate the 1-$\sigma$ confidence intervals 
and the solid lines show the median SNR 
$\langle \rho_{\text{tot}}\rangle$ for 300 realizations. 
The top panel of Fig.~\ref{fig:popa-growth} shows the results 
from fixing $R, \alpha_{\rm pow}, m_{\rm max}$ to their median
values. 
The bottom panel of Fig.~\ref{fig:popa-growth} is the same as the
top, except now $R$, $\alpha_{\rm pow}$ and $m_{\textrm{max}}$
are allowed to vary. 
We give the median value of the 
SNR $\langle \rho_{\text{tot}}\rangle$ and the 1-$\sigma$ 
confidence intervals for both analyses after five years in 
Table~\ref{tab:PopA}.

Figure~\ref{fig:popa-growth} and Table~\ref{tab:PopA} show that
after five years of detector operation time, the total SNR for
the GW memory in the population is approaching the SNR threshold 
of three.
Specifically, this threshold is close to the upper limit of the
1-$\sigma$ confidence interval for the conservative estimate 
(the red region) and the lower limit of the confidence interval
for the upper bound (the purple region).
The SNR for the memory effect does not differ greatly between the
populations with fixed and with varied parameters; the only 
obvious difference is a somewhat greater width of the 1-$\sigma$
confidence intervals when the population parameters are varied.
This is not surprising, because the fixed population does not
incorporate uncertainties on the rate, the maximum mass, and 
the power law, whereas the varied population does.

Because the results with and without the SNR cut for the higher
order modes are not very different, it is of interest to know
what fraction of the events pass this cut.
This is highlighted in Table~\ref{tab:SNRpercPopA}.
It shows that the majority of the events (around 70\%) do not 
pass this cut.
Thus, despite the large number of these events, their SNR is 
generally sufficiently small that they do not make a substantial
difference to the total SNR for the memory effect.
Table~\ref{tab:SNRpercPopA} also shows that it will be unlikely
for a given realization of a population to have an event in the
population that has $\rho^{\rm mem}_N \geq 1$.
Thus, the majority of the total SNR for the memory comes from the
louder subset of events that satsify the criteria 
$\rho^{\rm hom}_N \geq 2$ and $\rho^{\rm mem}_N < 1$.
This also was noted in~\cite{Lasky:2016knh}.

\begin{table}[htb]
    \centering
    \caption{Percentages of detections that satisfy the given criteria for the SNRs for the higher-order modes and for the memory effect for individual detections.
    The criteria are given for both the fixed and varied popluations dicussed in Sec.~\ref{sec:bbhpop}.}
    \begin{ruledtabular}
    \begin{tabular}{cccc}
       Population A  & $\rho^{\rm hom}_N < 2$ & $\rho^{\rm hom}_N \geq 2$  & $\rho^{\rm hom}_N \geq 2$ 
       \\
       & $\rho^{\rm mem}_N < 1$  & $\rho^{\rm mem}_N < 1$ & $\rho^{\rm mem}_N \geq 1$ \\
       \hline
       Fixed & $69.5\%$ & $30.4\%$ & $0.1\%$  \\
       Varied & $70.4\%$ & $29.5\%$ & $0.1\%$
         \\
    \end{tabular}
    \end{ruledtabular}
    \label{tab:SNRpercPopA}
\end{table}

\section{Discussion}
\label{sec:discussion}

In this paper, we investigated the prospects for detecting the
nonlinear GW memory effect by the advanced LIGO and Virgo detectors
in different populations of BBHs. 
We first noted that of three commonly used methods to compute
the memory effect, two produced similar results, whereas
the other one differed by a larger amount.
The two methods that more closely agreed made fewer approximations
to compute the GW memory waveforms, and thus seem to be the more
reliable waveform models for computing the GW memory effect
and performing estimates of when the memory effect will be
detected.

We also revisited the criteria used in~\cite{Lasky:2016knh} 
for assessing when an individual event will provide useful 
evidence for the presence of the memory in the population.
An important insight in~\cite{Lasky:2016knh} was that 
degeneracies in the GW mode $h_{22}$ lead to the SNR of 
the memory effect being uncertain for a single detection.
However, even if the SNR of the memory effect is small, 
as long as higher-order modes of the GWs 
are measurable for each individual BBH detection, then the
event will be useful for contributing to the total SNR for the 
memory effect in the population.
We performed Bayesian inference on several simulated BBH 
detections to find that on average, a network SNR of 2 in the
higher-order modes is sufficient to determine the memory effect's
sign (and thus its SNR for that event).
This criterion was similar to the one used in~\cite{Lasky:2016knh},
but it used a different subset and combination of the 
higher-order modes.

We then simulated two classes of populations of BBHs to determine
when the memory effect would be present in these populations.
We first looked at the population of GW150914-like BBHs that was
considered in~\cite{Lasky:2016knh}. 
Our results were consistent with those in~\cite{Lasky:2016knh} 
when we used the same GW waveform model as 
in~\cite{Lasky:2016knh}, but the SNR of the memory was notably
smaller when computed with the more recent waveform models that
make use of fewer approximations.
We then investigated the SNR for the memory effect in the simplest
model for the astrophysical population of BBHs that was inferred 
from the first ten GW detections of BBHs in~\cite{abbott2019binary}.
We considered two cases of this model, one where the parameters
of the mass distribution and the rate were fixed to the median
values, and one where we considered different realizations
of the mass distribution and rate.
In both cases, the SNR for the memory in the population was near 
the threshold of detection after five years (SNR of three), 
when using one of the more recent GW memory waveform models.
The spread of SNRs over different realizations of the populations
for the two cases was larger when the parameters were not fixed
to their median values, though, as a result of taking into 
account the additional uncertainties on the parameters 
describing the population.

\textit{Note}:
While this work was being completed, a preprint by H\"ubner 
\textit{et al.}~\cite{Hubner:2019sly} appeared that 
estimated the number of BBH observations required to detect
the nonlinear GW memory effect in BBH populations. 
There were several differences in methodology between this paper 
and~\cite{Hubner:2019sly}.
First,~\cite{Hubner:2019sly} computed evidence ratios for
signal hypotheses including and omitting the GW memory effect
and the Bayes factor (BF) for the presence versus the absence of
the memory effect in the population of BBHs (rather than computing
the total SNR for the memory effect, as was done in this paper).
Second, they used the higher-mode model rather than the
quadrupole model as the fiducial waveform model for the GW 
memory effect.
Third, they use a different model for the population of BBHs:
namely, model B of~\cite{abbott2019binary} for a specific
set of parameters given in~\cite{Hubner:2019sly}.
Fourth, they do not exclude events for which the sign of the
GW memory effect is not well determined.
With these differences in methodology, they find
$1830_{-1100}^{+1730}$ detections (errors are 90\% confidence 
intervals) are needed to reach a  $\log\mathrm{BF} = 8$ for
the GW memory effect. 

A direct comparison of our results will require 
additional future work.
As a rough comparison, we computed the number of detections
needed to reach a total memory SNR 
$\langle \rho_{\text{tot}}\rangle = 3$ for the same population
as in~\cite{Hubner:2019sly} using their same waveform model for the 
GW memory effect. 
We find we need $1488_{-879}^{+725}$ (errors are 90\% confidence 
intervals) to reach our SNR threshold.
Thus, the results seem roughly consistent.

\section*{Acknowledgements} 
We thank Yanbei Chen and Samaya Nissanke for their input in the
early stages of this work. 
We also thank Aaron Johnson and Paul Lasky for their correspondences 
about the different memory waveform models. Furthermore, we thank Marc Favata for helpful comments on the manuscript.
O.M.B.\ acknowledges funding from Vici research program 'ARGO' with project number 639.043.815, financed by the Dutch Research Council (NWO). 
D.A.N.\ acknowledges the support of the Netherlands Organization
for Scientific Research through the NWO VIDI Grant No.~639.042.612-Nissanke.
P.S.\ acknowledges NWO Veni Grant No. 680-47-460. This paper has LIGO document number P2000013.

\appendix

\section{Quadrupole ``kludge'' memory waveform model}
\label{sec:appquadkludge}

In this appendix, we discuss one additional waveform model 
that was recently used in~\cite{Johnson:2018xly} to make estimates 
of the SNR for the GW memory effect in a wide range of GW detectors.
We show that it produces a signal related to the GW memory effect
that is roughly half the amplitude of the curves shown in 
Fig.~\ref{fig:mem_models}, and which will also have some small oscillatory part that would typically not be expected
in the corresponding spherical-harmonic modes for the memory.

The model of~\cite{Johnson:2018xly} begins with the procedure
in~\cite{Thorne:1992sdb}, which proposed a method to 
simplify evaluating the angular integrals that appear in
Eq.~\eqref{eq:memcor}.
The method is to compute the GW polarizations [similarly to what
was done in Eq.~\eqref{eq:hmemPols}], but to work in coordinates
adapted to the detector and the incoming radiation rather than 
the source.
These coordinates are defined by choosing as the $x$ direction any 
direction that is transverse to the vector pointing between
the detector and the source.
The GW polarizations are then computed with respect to the complex combination 
$e_+^{ij} + i e_\times^{ij}$ of polarization tensors, where
$e_+^{ij} = (\hat x^i \hat x^j - \hat y^i \hat y^j)/2$ and 
$e_\times^{ij} = (\hat x^i \hat y^j + \hat y^i \hat x^j)/2$, and
where $\hat x^i$ and $\hat y^i$ are unit vectors in the $x$ and
$y$ directions, respectively, in the frame described 
in~\cite{Thorne:1992sdb}.
This simplifies the part of the integral proportional to
$n'_j n'_k/(1-\mathbf n' \cdot \mathbf n)$ [although potentially
at the expense of complicating the expansion of $dE/(d\Omega'du)$, 
which we had previously been computing in terms of multipole
moments of the GW strain in coordinates in which the binary
is in the $x$-$y$ plane].

Johnson \textit{et al.}~\cite{Johnson:2018xly} 
compute the GW polarizations following~\cite{Thorne:1992sdb}. 
Rather than working out the detailed transformation of the 
multipole expansion of the luminosity per solid angle
between the source coordinates and their coordinates for each line 
of sight from source to detector, they make the following 
approximate model that they describe as a ``kludge'':
They take the angular dependence of the memory given in
Eq.~\eqref{eq:mem} [i.e., $\sin^2\iota (17+\cos^2\iota)$],
but instead of multiplying by the integral of $|\dot h_{22}|$
as in Eq.~\eqref{eq:mem}, they multiply by the integral
of $\dot h_+^2$ evaluated at an inclination of $\iota=0$
in the \textit{source} coordinates, where $h_+$ is the full plus
polarization, including (in principle) all $(\ell,m)$ modes
[as in Eq.~\eqref{eq:hlm}].
See~\cite{Johnson:2018xly} for the details about the rationale 
behind this prescription.
This procedure leads to a real GW strain, which in the coordinates 
of~\cite{Thorne:1992sdb} implies that the GW memory strain is 
plus polarized and is given by
\begin{equation} \label{eq:memJplus}
    h^{\mathrm{(K)}}_{\mathrm{mem},+}(u) = \frac{r}{68\pi} \Phi(\iota)
    \int\limits_{-\infty}^{u} \dot{h}_{+}^2|_{\iota=0}\, du' \, .
\end{equation}
We have defined $\Phi(\iota) = \sin^2(\iota)(17+\cos^2\iota)$
for convenience.

For the inclination and phase $\phi_c$ that points to the line of 
sight to the detector, the polarizations $e_+$ and $e_\times$ 
defined in the source coordinates in Eq.~\eqref{eq:eplus_cross} are 
transverse (and traceless) tensors with respect to the direction
of the line of sight.
Thus, these polarizations in the source coordinates and those in 
the coordinates of~\cite{Thorne:1992sdb} must be related by a
rotation about the line of sight between the source and detector.
In the quadrupole approximation in Sec.~\ref{sec:memmodels},
the memory is plus polarized, but it is also plus polarized 
in Eq.~\eqref{eq:memJplus}; thus, at this level of approximation 
for computing the GW polarizations associated with the GW memory
effect, the rotation is trivial and the two sets of polarizations
are equivalent.
 
Let us then write the integral in Eq.~\eqref{eq:memJplus} using 
the quadrupole approximation that only $h_{22}$ contributes
to $\dot h_+$ in the integral in Eq.~\eqref{eq:memJplus}. 
We will denote this further approximation by
$h^{\mathrm{(QK)}}_{\mathrm{mem},+}(u)$.
A straightforward calculation shows that the memory computed via 
the kludge method of~\cite{Johnson:2018xly} relates to the
plus polarization of the GW memory effect in the
quadrupole approximation in Eq.~\eqref{eq:mem} as follows:
\begin{equation}
\begin{split}
    h^{\mathrm{(QK)}}_{\mathrm{mem},+}(u) = &
    \frac{24}{17} |^{(-2)}Y_{22}(0,0)|^2 h^\mathrm{mem}_+(u) \\
    & + \frac{r}{136\pi}\Phi(\iota)
    \int\limits_{-\infty}^{u} \, du'
    \Re\{[\dot h_{22} {}^{(-2)}Y_{22}(0,\phi_c)]^2\} \\
    & \approx 0.56 h^\mathrm{mem}_+(u) \\
    & + \frac{r}{136\pi}\Phi(\iota) \int\limits_{-\infty}^{u} \, du'
    \Re\{[\dot h_{22} {}^{(-2)}Y_{22}(0,\phi_c)]^2\} \, .
\end{split}
\end{equation}
The second term involving the integral of the square of the
real part of $\dot h_{22}$ will generally be small 
(see, e.g.,~\cite{Favata:2008yd}), and will oscillate at
twice the frequency of the mode $h_{22}$ (this is likely
the origin of the oscillations in the memory waveform
model in~\cite{Johnson:2018xly}).
Thus, the quadrupole approximation to the procedure 
in~\cite{Johnson:2018xly} will typically produce a waveform
that is roughly half the amplitude of the two models that
use fewer approximations in Sec.~\ref{sec:memmodels}, and it
will contain an additional unexpected oscillatory part.
As a result, we do not include it in our comparison in 
Sec.~\ref{sec:memmodels}.

\section{Analogy based on stacking memory signals}
\label{sec:appmemstack}

\begin{figure}[htb]
  \centering
  \subfigure{\includegraphics[width=0.99\columnwidth]{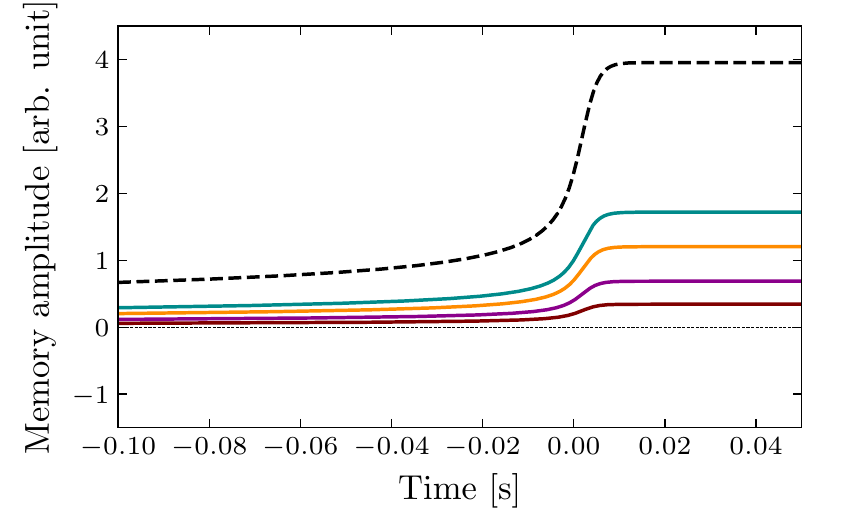}} 
  \subfigure{\includegraphics[width=0.99\columnwidth]{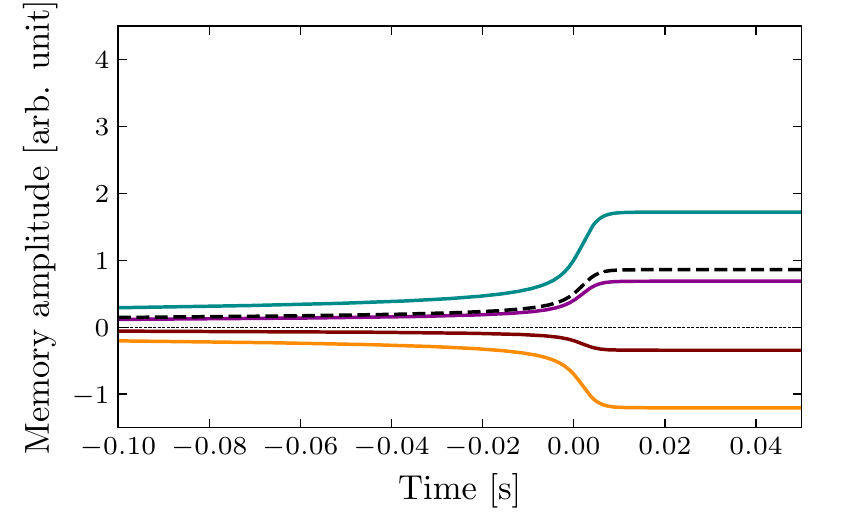}}
    \caption{Schematic illustration of why it is necesary to know
    the sign of the GW memory effect. The solid cyan, orange, 
    purple, and brown curves are four GW memory waveforms. 
    The dashed black curve is the sum of the waveforms. 
    \textit{Top}: We assume we know the sign of the GW memory 
    for all events, so that when we add the individual events
    together, the combined signal is roughly four times larger
    than the average size of the individual signals.
    \textit{Bottom}: We assume we do not know the sign of the
    GW memory, so that when we add the different waveforms,
    the ones with opposite sign cancel and the combined signal
    is on the same order as the individual waveforms.}
    \label{fig:stack_ill}
\end{figure}

There is a simple analogy one can make to describe why it is
important to know the sign of the GW memory effect to compute
the total SNR in a population of BBHs.
This is illustrated in Fig.~\ref{fig:stack_ill} for four
memory signals.
It shows in the top panel that if the sign of the memory signals
are known, then when $N_e$ memory signals are added together the 
net signal will be roughly $N_e$ times the individual signals,
assuming the signals are on roughly the same size.
The bottom panel shows that this does not occur when the GW
memory waveforms are added with different signs.
When $N_e$ realizations of independent Gaussian noise are 
added together, the variance grows like $N_e$.
Thus, the SNR grows like $\sqrt{(N_e)^2/N_e} = \sqrt{N_e}$
when the signals are added with the same sign, but it exhibits
much slower (if any growth) with $N_e$ if they are added with
random signs.

\section{A second example of determining the sign of the memory
effect}
\label{sec:appmemsign}

\begin{figure*}[htb]
  \centering
  \subfigure{\includegraphics[width=0.45\textwidth]{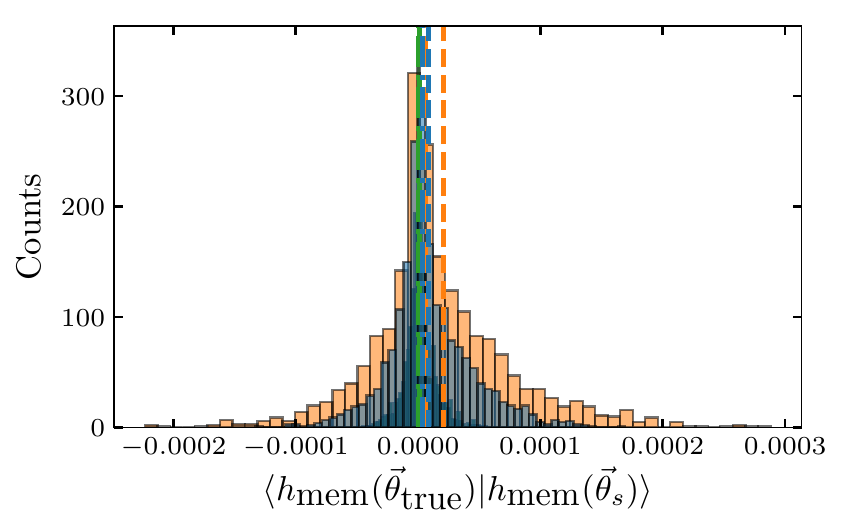}}\hfill
  \subfigure{\includegraphics[width=0.45\textwidth]{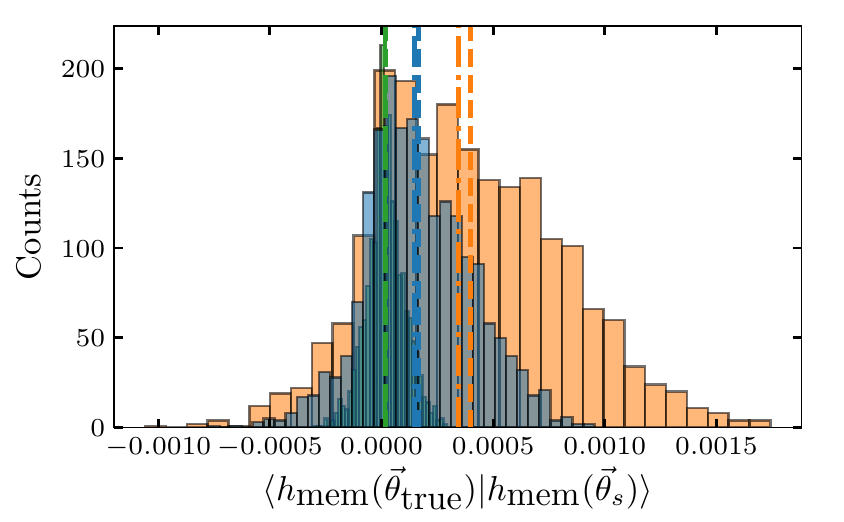}}\hfill
  \subfigure{\includegraphics[width=0.45\textwidth]{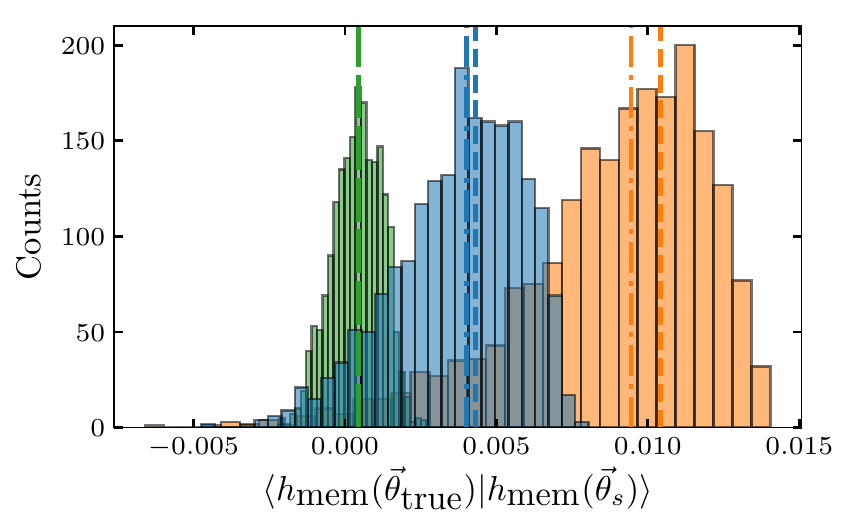}}\hfill
  \subfigure{\includegraphics[width=0.45\textwidth]{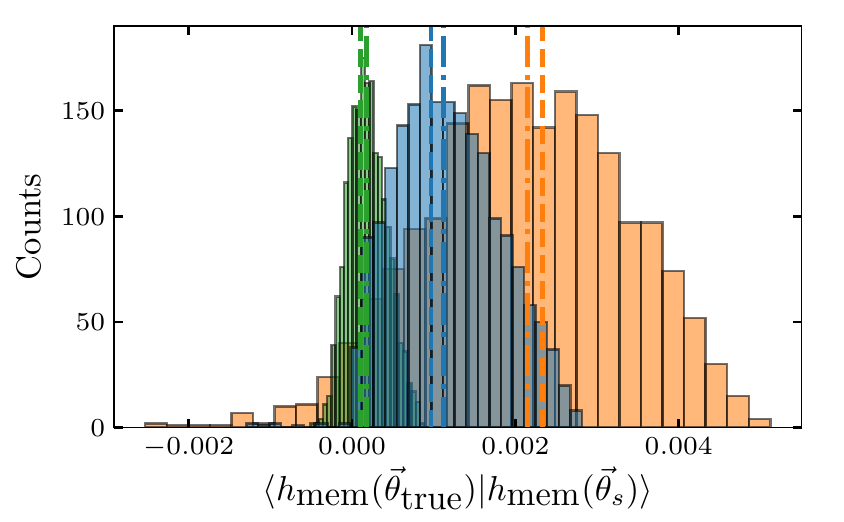}}\hfill
    \caption{Distribution of noise-weighted inner products for the
    memory effect for a nonspinning binary with 
    $m_1= 30 M_{\odot}$, $m_2= 20 M_\odot$ at a fixed luminosity distance $d_{L} = 500$ Mpc with sky location and polarization given by $\alpha = 2.3$ rad, $\delta = 1.0$ rad and $\psi = 0.3$ rad. 
    As in Fig.~\ref{fig:mem_overlap_3030}, this figure shows
    the inner product of the true signal with template waveforms that are consistent with the posterior PDFs: 
    $\langle h_{\textrm{mem}} (\vec{\theta}_{\textrm{true}})| 
    h_{\textrm{mem}} (\vec{\theta}_s)\rangle $. 
    Advanced LIGO Hanford, LIGO Livingston and Virgo are shown in green, blue and orange, respectively. 
    The vertical dashed lines represent the injected values of the optimal SNR squared, and the vertical dotted-dashed lines represent the median values of the distributions.
    Here the SNR in the higher-order modes  
    $\rho^{\mathrm{hom}}_{N}$ is approximately 1, 2, 3, and 4 
    going clockwise from the top left.
    These different SNRs were achieved just by varying the 
    inclination angle of the source (the specific values are
    given in the text).}
    \label{fig:mem_overlap_3020}
\end{figure*}

Here, we illustrate an example of a binary for which the sign
of the GW memory effect is more challenging to measure than
in the more typical example in Sec.~\ref{sec:signmeasurement}. 
We perform Bayesian inference as described in Sec.~\ref{sec:PE} 
on a $m_1 =30 M_{\odot}$, $m_2=20 M_{\odot}$ binary at a fixed
luminosity distance $d_L = 500$ Mpc. 
Here, we now vary $\rho^{\mathrm{hom}}_N$ by changing the 
inclination $\iota$ rather than the luminosity distance $d_L$. 
In Fig.~\ref{fig:mem_overlap_3020} (going clockwise from the
top left) are the distributions of the inner product 
$\langle h_{\textrm{mem}} (\vec{\theta}_{\textrm{true}})| 
h_{\textrm{mem}} (\vec{\theta}_s)\rangle $ 
for $\iota = 3.0$, 0.3, 0.48, and 2.4 (where the SNRs 
$\rho^{\mathrm{hom}}_{N}$ are given by 1, 2, 3, and 4).
The oscillatory SNR for the three detector network is above
the threshold for detection for all four inclination angles. 
The histograms for advanced Virgo, LIGO Hanford, and LIGO
Livingston are shown in orange, green, and blue, respectively.

When there is an SNR of 2 in the higher-order modes, there
is more support for the true sign of the memory effect in the
Virgo detector than in the two LIGO detectors.
This occurs for the following reasons:
First, the LIGO antenna patterns are not very sensitive to
the plus polarization for the sky location and polarization
of the binary.
Furthermore, the source is located almost directly above
the plane formed by the three detectors, and there is an
approximate degeneracy between the true location of the
source and the source on the opposite side of the sky.
Thus, there is an additional degeneracy between the sky
location and polarization in addition to the degeneracy between the polarization and phase $\phi_c$.
These facts combine to require a slightly higher SNR
of closer to 3 in the higher-order modes before the
sign of the memory is more confidently measured by Virgo
and LIGO Livingston (LIGO Hanford is not sensitive to 
the plus polarization of the binary).
This case is somewhat unusual, because of the very specific
sky location and polarization leading to a poor sensitivity
to the GW memory in the LIGO detectors; the results in 
Fig.~\ref{fig:mem_overlap_3030} are more representative
of most binaries that we simulated.

\bibliography{citations.bib}

\end{document}